\documentclass[fleqn,twoside]{article}
\usepackage[headings]{espcrc2}

\usepackage{graphicx}
\usepackage{epsfig}
\usepackage{amssymb}
\usepackage[figuresright]{rotating}

\newcommand{\AmS}{{\protect\the\textfont2
  A\kern-.1667em\lower.5ex\hbox{M}\kern-.125emS}}

\def\q2{Q^2}
\def\ele{e^+e^-}
\def\pp{p\bar p}

\def\ii{\'\i}
\def\oo{\'o}

\def\etjet{E_T^{\rm jet}}
\def\etajet{\eta^{\rm jet}}
\def\kt{k_T}
\def\yc{y_{\rm cut}}
\def\ns{n_{\rm sbj}}

\def\oass{{\cal O}(\as^2)}
\def\as{\alpha_s}
\def\ps{\psi(r)}
\def\xo{x_{\gamma}^{\rm obs}}
\def\etajb{\eta^{\rm jet}_{\rm B}}
\def\etjb{E^{\rm jet}_{T,{\rm B}}}
\def\asz{\as(\mz)}
\def\as{\alpha_s}
\def\mz{M_Z}
\def\asmz#1#2#3#4#5#6{\asz = #1\pm #2\ {\rm (stat.)}\ ^{+#4}_{-#3}\ {\rm (exp.)}\ ^{+#6}_{-#5}\ {\rm (th.)}}
\def\asbj{\alpha^{\rm sbj}}
\def\etsbj{E_T^{\rm sbj}}
\def\etasbj{\eta^{\rm sbj}}
\def\phisbj{\phi^{\rm sbj}}
\def\phijet{\phi^{\rm jet}}
\def\qq{q\bar q}
\def\Journal#1#2#3#4{{#1} {#2} (#3) #4}

\def\NPB{{\em Nucl. Phys.} {\bf B}}
\def\PLB{{\em Phys. Lett.}  {\bf B}}
\def\PRL{{\em Phys. Rev. Lett.}}
\def\PRD{{\em Phys. Rev.} {\bf D}}

\def\ZPC{{\em Z. Phys.} {\bf C}}

\def\EPC{{\em Eur. Phys. Jour.} {\bf C}}

\def\CPC{{\em Comp. Phys. Comm.}}

\title{Jet substructure at HERA\thanks{Talk given in the
    ``Ringberg Workshop: New Trends in HERA Physics 2008'', Ringberg
    Castle, Tegernsee, Germany, October 2008.}}

\author{Claudia Glasman\address[MCSD]{Universidad Aut\oo noma de Madrid, \\
    Departamento de F\ii sica Te\oo rica, Facultad de Ciencias\\
    Cantoblanco, E-28049 Madrid, Spain}
({\it On behalf of the H1 and ZEUS Collaborations})}

\runtitle{Jet substructure at HERA}
\runauthor{C. Glasman}

\begin{document}

\begin{abstract}
A review is presented of jet substructure measurements from the H1 and
ZEUS Collaborations at HERA. The results presented include tests of
perturbative QCD, comparison of the properties of quark and gluon
jets, the comparison of the pattern of QCD radiation in different hard
scattering processes, determination of the strong coupling and
studies of the underlying subprocesses dynamics and of the pattern of
parton radiation.
\vspace{1pc}
\end{abstract}

\maketitle

\section{Introduction}

Perturbative calculations deal only with partons in the final
state. However, hadrons and not partons are observed in the
detector. The hadrons are the result of the fragmentation of the
partons such that all the hadrons originating from a given parton are
contained in a narrow region around the direction of the original
parton, what is commonly called a jet. Therefore, the first step in
comparing data and theory is to reconstruct the parton topology. This
is best done with jet algorithms, which define the jets in a very
specific way. The jets obtained display the following characteristics:
{\it (i)} the longitudinal momentum ($p_L$) distribution of the
hadrons in a jet should scale with the jet energy; {\it (ii)} the
transverse momentum ($p_T$) distribution of the hadrons should have a
mean value of about 300 MeV. Thus, the mean angle ($\sim p_T/p_L$)
between a hadron and the jet axis and the size of a cone which
contains a constant fraction of energy will decrease with increasing
jet energy. At high energies, this picture is modified since gluon
emission from the original parton overcomes the fragmentation effects
and so the structure of the jets is driven by radiation, which has the
advantage of being calculable in perturbative QCD (pQCD).

Thus, the investigation of the internal structure of jets gives
insight into the transition between the parton produced in the hard
process and the experimentally observed jet of hadrons. QCD predicts
that {\it (i)} the jet substructure is driven by gluon emission off
primary partons at sufficiently high jet transverse energy, $\etjet$,
where fragmentation effects become negligible, {\it (ii)} jets
originating from gluons are broader than those originating from quarks
since the gluon-initiated jets radiate more due to their larger colour
charge and {\it (iii)} the jet substructure depends mainly on the type
of primary parton (quark or gluon) from which it originated and to a
lesser extent on the hard scattering process. The experimental results
from HERA which confirm these predictions are presented in this report.

The jet substructure can be studied in terms of two different
observables, namely the jet shape~\cite{prl:69:3615} and the subjet
multiplicity~\cite{np:b383:419}. The jet shape is defined as the
fraction of the jet transverse energy that lies inside a cone in the
$\eta-\phi$ plane of radius $r$, $E_T(r)$, concentric with the jet
axis. The mean integrated jet shape is given by

$$\langle\psi(r)\rangle=\frac{1}{N_{\rm jets}}\sum_{\rm jets}\frac{E_T(r)}{\etjet}.$$
To compute the jet shape, the jets can be reconstructed either with
the cone~\cite{pr:d45:1448} or the $\kt$~\cite{np:b406:187}
algorithms. In contrast, the subjet multiplicity can be computed using
only the jets reconstructed with the $\kt$ algorithm and is obtained
by reapplying the algorithm until for every pair of particles inside
the jet, the quantity

$$d_{ij}={\rm min}(E_{Ti},E_{Tj})^2[(\eta_i-\eta_j)^2+(\phi_i-\phi_j)^2]$$
is above $\yc\cdot (\etjet)^2$, where $\yc$ is the resolution
parameter. The mean subjet multiplicity is given by 

$$\langle\ns(\yc)\rangle=\frac{1}{N_{\rm jets}}\sum_{i=1}^{N_{\rm jets}} \ns^i(\yc).$$

The calculations of the jet substructure using QCD-based leading-order
(LO) Monte Carlo (MC) models, such as {\sc Pythia}~\cite{cpc:46:43},
{\sc Herwig}~\cite{cpc:67:465}, {\sc Ariadne}~\cite{cpc:71:15} and
{\sc Lepto}~\cite{cpc:101:108}, are approximations based on a
parton-shower approach. In fixed-order QCD calculations, the lowest
order provides no structure for the jets since there is only one
parton inside a jet. Higher-order terms give the lowest non-trivial
contribution to the jet substructure. In particular, it is possible to
obtain next-to-leading-order (NLO) calculations for jet substructure
in neutral current (NC) deep inelastic scattering (DIS) in the
laboratory frame from $\oass$ predictions since, in this case, three
partons can be inside one jet. Thus, measurements of jet substructure
provide a stringent test of pQCD directly beyond LO. The pQCD
calculations of the jet shape and subjet multiplicity are obtained via

$\langle 1-\ps\rangle=$
$$\frac{\int_r^R dE_T \; (E_T/\etjet) [d\sigma(ep \rightarrow 2 {\rm partons})/dE_T]}{\sigma_{\rm jet}(\etjet)}$$

$\langle\ns(\yc)\rangle=$
$$1+\frac{1}{\sigma_{\rm jet}}\sum_{j=2}^{\infty}(j-1)\cdot\sigma_{{\rm sbj},j}(\yc)=1 + C_1 \ \as + C_2 \ \as^2.$$

\section{Tests of pQCD using jet substructure}

At HERA, the main source of jets is photoproduction. At LO QCD, two
processes contribute to the jet production cross section, resolved, in
which the photon interacts via its partonic structure, and direct
processes, in which the photon interacts as a point-like
particle. Both processes give rise to two jets in the final state. The
observable 
$\xo\equiv (1/E_{\gamma})(\sum_{\rm jets}\etjet e^{-\etajet})$
measures the energy invested by the photon in producing the dijet
system and can be used to separate resolved and direct processes since
they populate different regions of phase space. Resolved processes
give rise to jets of quarks and gluons in the final state. On the
other hand, direct processes are dominated by quark jets, so the
$\etajet$ dependence of the jet substructure is expected to show
quark-like jets in the rear direction and gluon-like jets in the
proton direction due to HERA dynamics, since for the dominant resolved
diagram ($g_pq_{\gamma}\rightarrow gq$), the gluon goes forward. On the
other hand, the jets in an inclusive-jet sample of NC or charged
current (CC) DIS processes are dominated by quark jets so that no
significant dependence with $\etajet$ is expected. Only for a dijet
sample, which contains a larger fraction of gluon jets, some
dependence with $\etajet$ could be observed.

\begin{figure}[h]
\setlength{\unitlength}{1.0cm}
\begin{picture} (18.0,5.0)
\put (0.0,-1.0){\centerline{\epsfig{figure=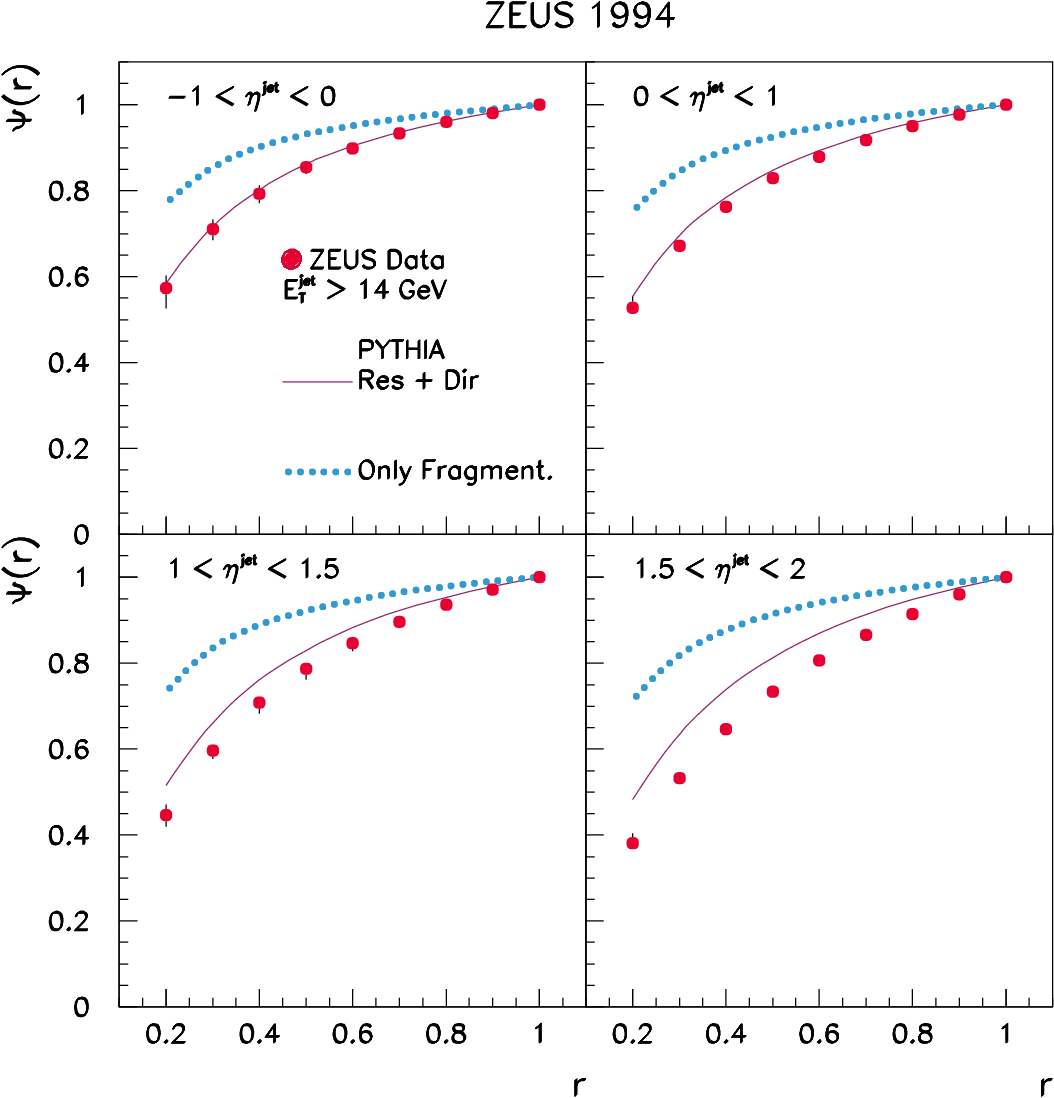,width=6cm}}}
\end{picture}
\caption{\label{fig1}
{Mean integrated jet shape in inclusive-jet photoproduction.}}
\end{figure}

The mean integrated jet shape was measured~\cite{epj:c2:61} for
inclusive-jet photoproduction using the iterative cone
algorithm with cone radius $R=1$. Figure~\ref{fig1} shows the mean
integrated jet shape as a function of $r$ for different regions of
$\etajet$. The jets become broader as $\etajet$ increases. The
comparison with the QCD-based MC calculations shows that models with
only fragmentation predict jets that are too narrow compared to the
data. In contrast, models which include initial- and final-state QCD
radiation describe the data well for $-1<\etajet<1$. Therefore, parton
radiation is the dominant mechanism responsible for the jet shape.

%Figure 1

\begin{figure}[h]
\setlength{\unitlength}{1.0cm}
\begin{picture} (18.0,5.0)
\put (0.0,-1.0){\centerline{\epsfig{figure=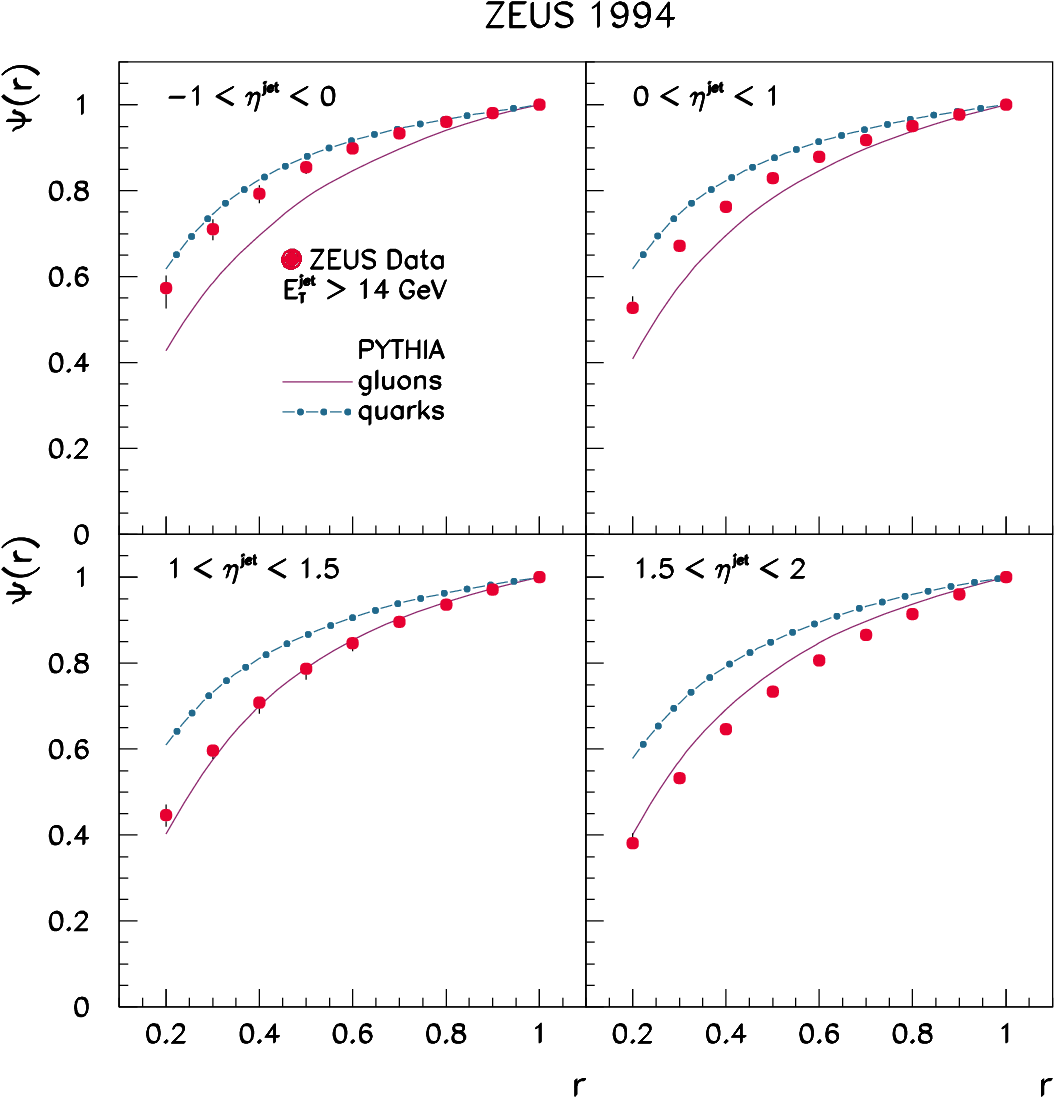,width=6cm}}}
\end{picture}
\caption{\label{fig2}
{Mean integrated jet shape in inclusive-jet photoproduction.}}
\end{figure}

The comparison of the measurements with predictions for samples of pure
quark- or gluon-initiated jets (see Fig.~\ref{fig2}) shows that the
measured jets are quark-like for $-1<\etajet<0$ and become
increasingly gluon-like as $\etajet$ increases, as predicted by QCD.

%Figure 2

\begin{figure}[h]
\setlength{\unitlength}{1.0cm}
\begin{picture} (18.0,4.5)
\put (0.0,-1.2){\centerline{\epsfig{figure=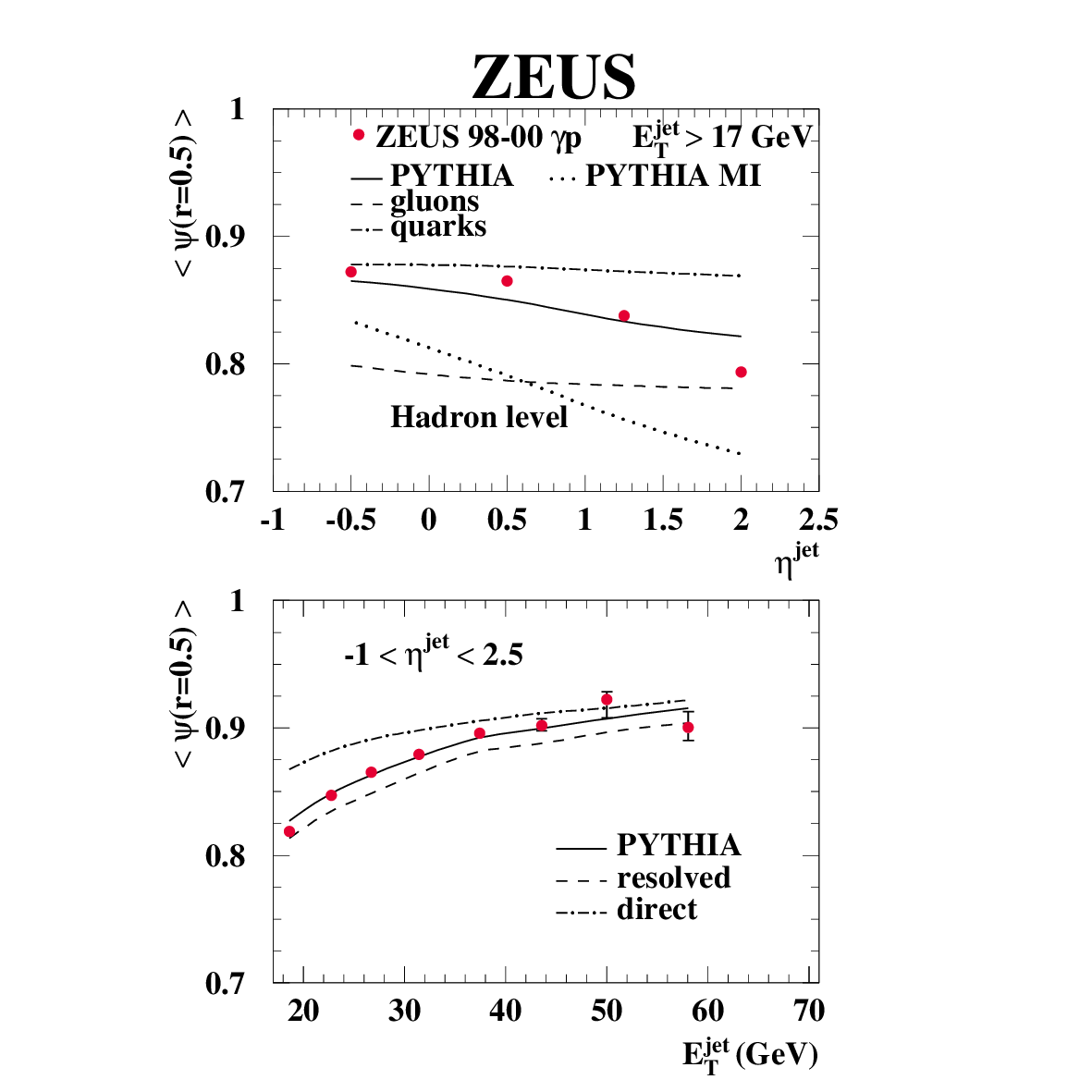,width=6.5cm}}}
\end{picture}
\caption{\label{fig3}
{Mean integrated jet shape in inclusive-jet photoproduction.}}
\end{figure}

To see more clearly the dependence of the jet shape with $\etajet$ and
$\etjet$, the mean integrated jet shape as a function of $\etajet$ and
$\etjet$ for a fixed value of $r=0.5$~\cite{np:b700:3} is presented in
Fig.~\ref{fig3}. This analysis was done using the $\kt$ algorithm to
reconstruct the jets. The measured jet shape decreases as $\etajet$
increases, which indicates that the jets become broader, and the
measured jet shape increases as $\etjet$ increases, which indicates
that the jets become narrower as $\etjet$ increases. The comparison
with the predictions for quark and gluon jets shows that the
broadening of the jets as $\etajet$ increases is consistent with an
increasing fraction of gluon jets. The comparison with the predictions
for resolved and direct processes shows that the events are dominated
by resolved processes at low $\etjet$, whereas the direct contribution
becomes increasingly more important as $\etjet$ increases.

%Figure 3

\begin{figure}[h]
\setlength{\unitlength}{1.0cm}
\begin{picture} (18.0,5.5)
\put (0.5,1.0){\epsfig{figure=dijets_det2_xg1.ps,width=6cm}}
\put (4.0,1.0){\epsfig{figure=dijets_det2_xg2.ps,width=6cm}}
\put (1.0,-1.0){\epsfig{figure=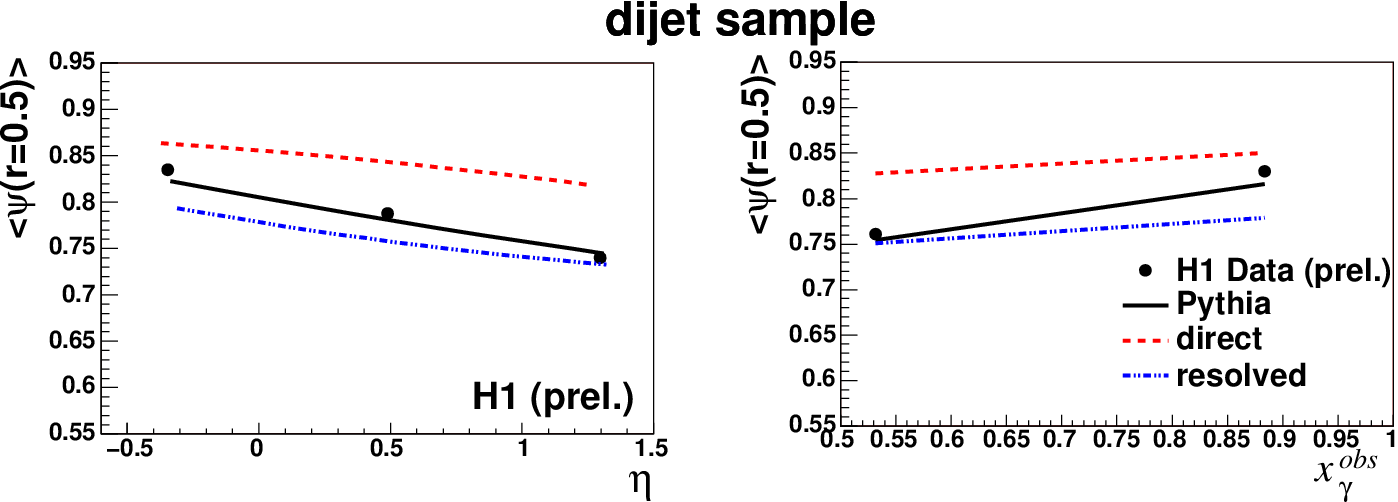,width=6cm}}
\end{picture}
\caption{\label{fig4}
{Mean integrated jet shape in dijet photoproduction.}}
\end{figure}

This last result was studied in more detail~\cite{h1-prelim-05-077} by
separating a dijet photoproduction sample in $\xo<0.75$, dominated by
resolved processes, and $\xo>0.75$, dominated by direct processes. The
jets in the resolved-enriched sample are broader than those for the
direct-enriched sample (see Fig.~\ref{fig4}). The QCD-based MC
predictions give a good description of the data and show that the data
is consistent with being dominated by resolved (direct) processes for
$\xo<0.75$ ($\xo>0.75$).

%Figure 4

\begin{figure}[h]
\setlength{\unitlength}{1.0cm}
\begin{picture} (18.0,1.9)
\put (-0.5,-1.0){\epsfig{figure=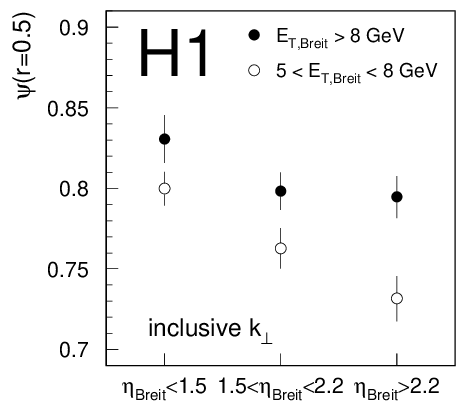,width=4cm}}
\put (3.5,-1.0){\epsfig{figure=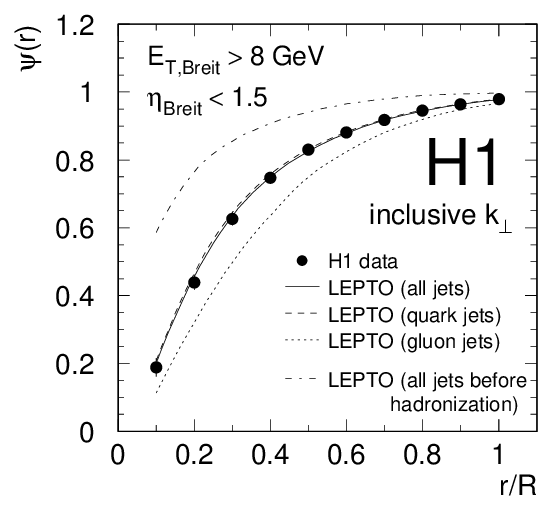,width=4cm}}
\end{picture}
\caption{\label{fig5}
{Mean integrated jet shape in dijet NC DIS in the Breit frame.}}
\end{figure}

The mean integrated jet shape was also measured in NC DIS in the Breit
frame for different regions of jet pseudorapidity, $\etajb$, and
transverse energy, $\etjb$, for a sample of dijet
events~\cite{np:b545:3}. The measurements show that the jet shape for
a fixed value of $r=0.5$ decreases as a function of $\etajb$. The
effect is more pronounced at low $\etjb$ (see Fig.~\ref{fig5}). The
data are well described by the predictions of the QCD-based MC models;
this comparison shows that the observed jet substructure is compatible
with that of quark-initiated jets. 

%Figure 5

\begin{figure}[h]
\setlength{\unitlength}{1.0cm}
\begin{picture} (18.0,4.5)
\put (1.0,-1.0){\epsfig{figure=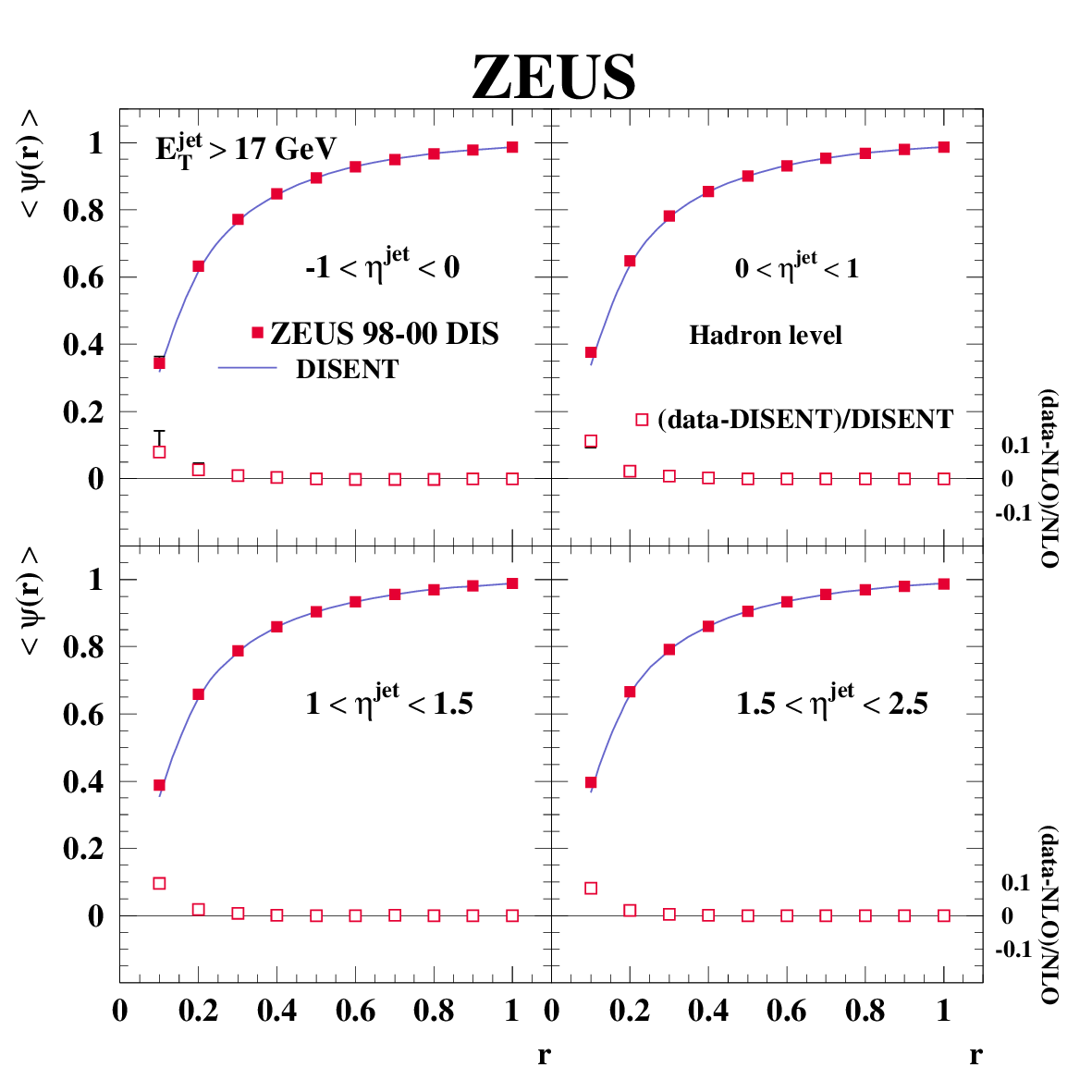,width=6cm}}
\end{picture}
\caption{\label{fig6}
{Mean integrated jet shape in inclusive-jet NC DIS in the laboratory
  frame.}}
\end{figure}

\begin{figure}[h]
\setlength{\unitlength}{1.0cm}
\begin{picture} (18.0,4.0)
\put (1.0,-1.0){\epsfig{figure=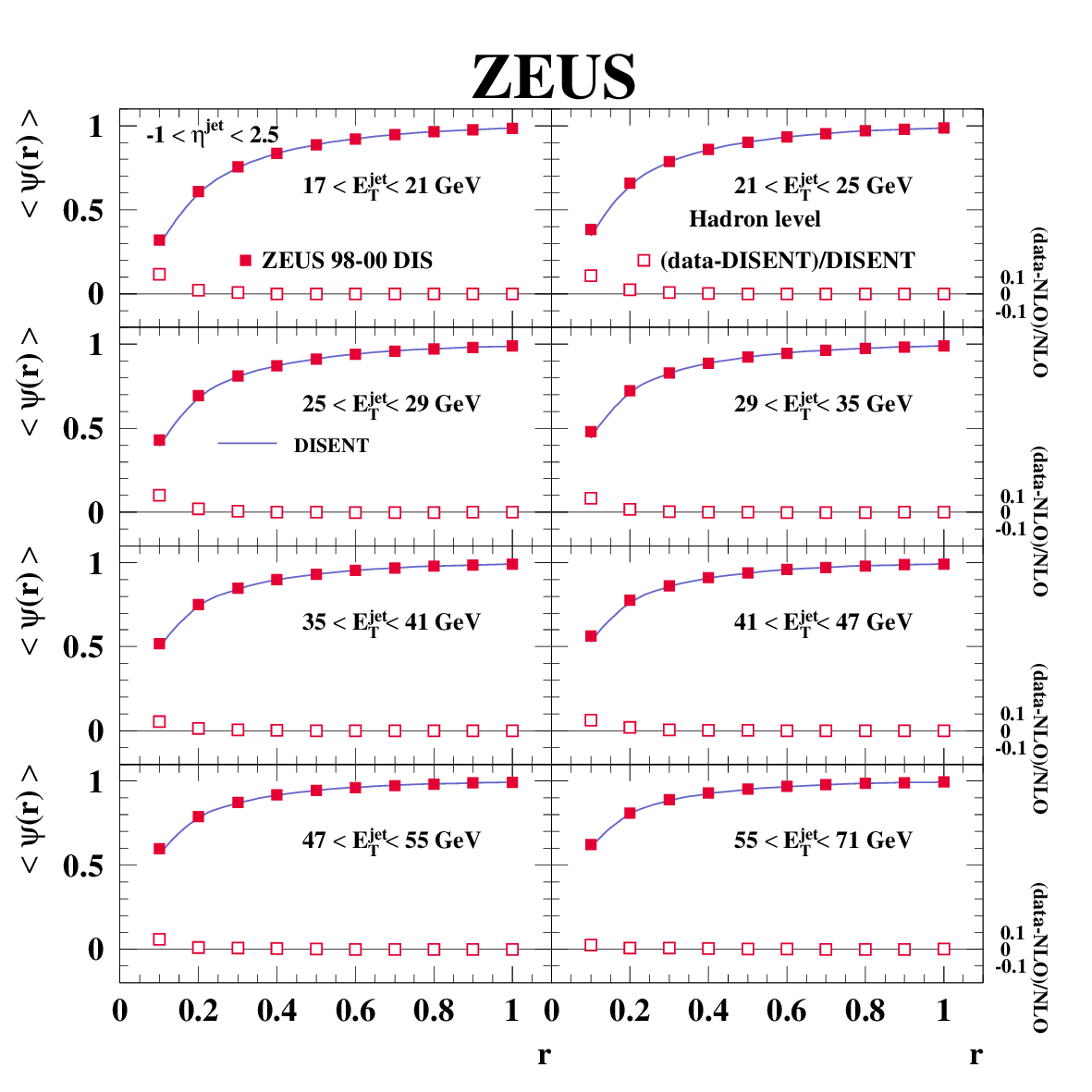,width=6cm}}
\end{picture}
\caption{\label{fig7}
{Mean integrated jet shape in inclusive-jet NC DIS in the laboratory
  frame.}}
\end{figure}

\begin{figure}[h]
\setlength{\unitlength}{1.0cm}
\begin{picture} (18.0,4.5)
\put (0.5,-1.0){\epsfig{figure=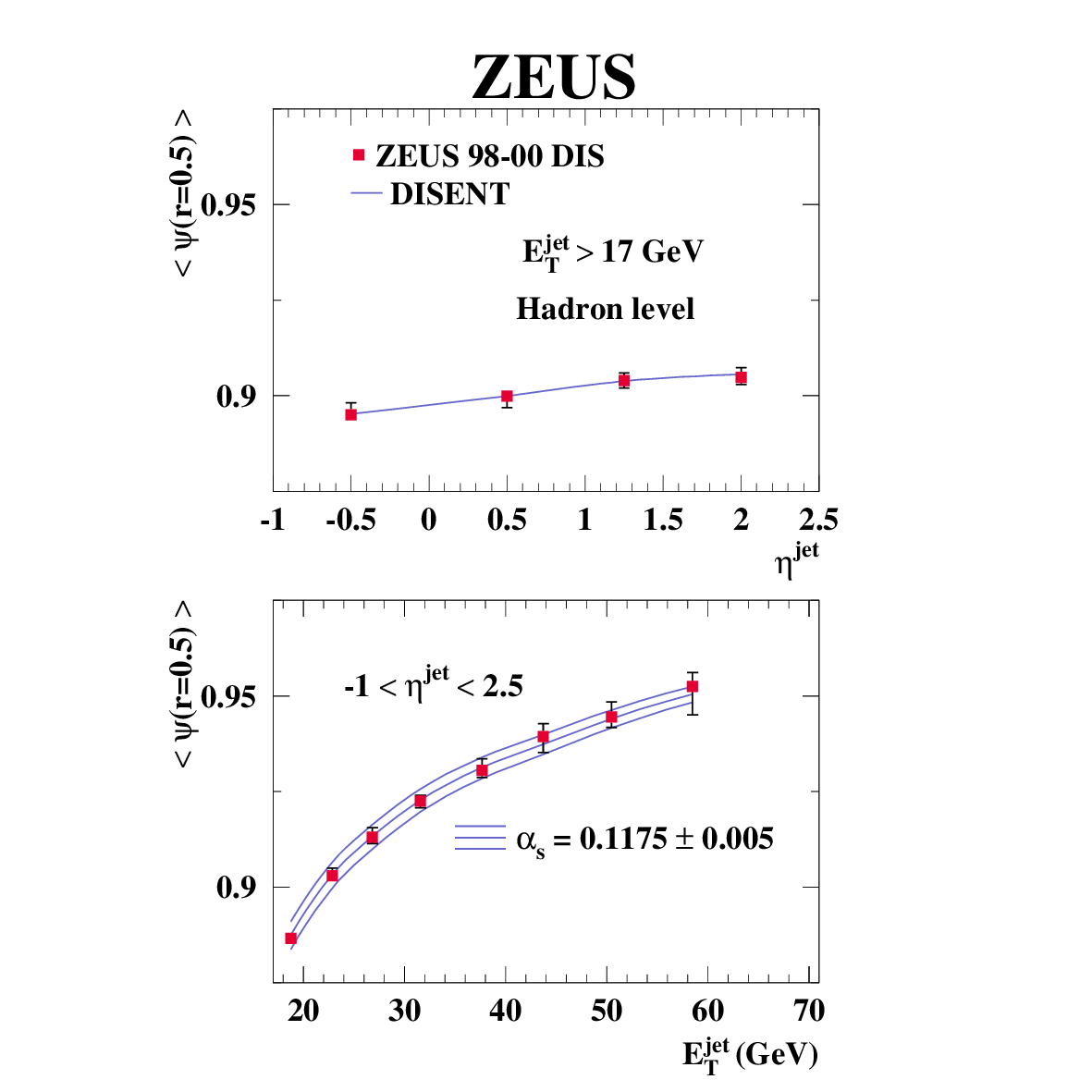,width=6cm}}
\end{picture}
\caption{\label{fig8}
{Mean integrated jet shape in inclusive-jet NC DIS in the laboratory
  frame.}}
\end{figure}

Figures~\ref{fig6} and \ref{fig7} show the integrated jet shape
measured in NC DIS in the laboratory frame for a sample of inclusive
jets as a function of $r$ in different regions of $\etajet$ and
$\etjet$~\cite{np:b700:3}. The measurements are compared to NLO
calculations~\cite{np:b485:291}, which give a good description of the
data for $r>0.1$. The $\etajet$ and $\etjet$ dependence of the mean
integrated jet shape was measured for a fixed value of $r=0.5$ (see
Fig.~\ref{fig8}). No significant variation with $\etajet$ is observed
in this case, consistent with the sample being dominated by quark jets
in all the phase space measured. It is observed that the jets become
narrower as $\etjet$ increases. The NLO calculations give a good
description of the data and show a sensitivity to the value of
$\as$. From the measured jet shape for $\etjet>21$~GeV, a value of
$\asz$ was extracted,
$\asmz{0.1176}{0.0009}{0.0026}{0.0009}{0.0072}{0.0091}$, which is
consistent with other determinations from more inclusive channels and
has a small experimental uncertainty. The theoretical uncertainties
are dominant, especially that arising from higher orders; a better
determination of $\as$ from this observable would be possible once
higher-order corrections for NC DIS become available.

%Figure 6

%Figure 7

%Figure 8

\begin{figure}[h]
\setlength{\unitlength}{1.0cm}
\begin{picture} (18.0,6.5)
\put (-0.5,2.5){\epsfig{figure=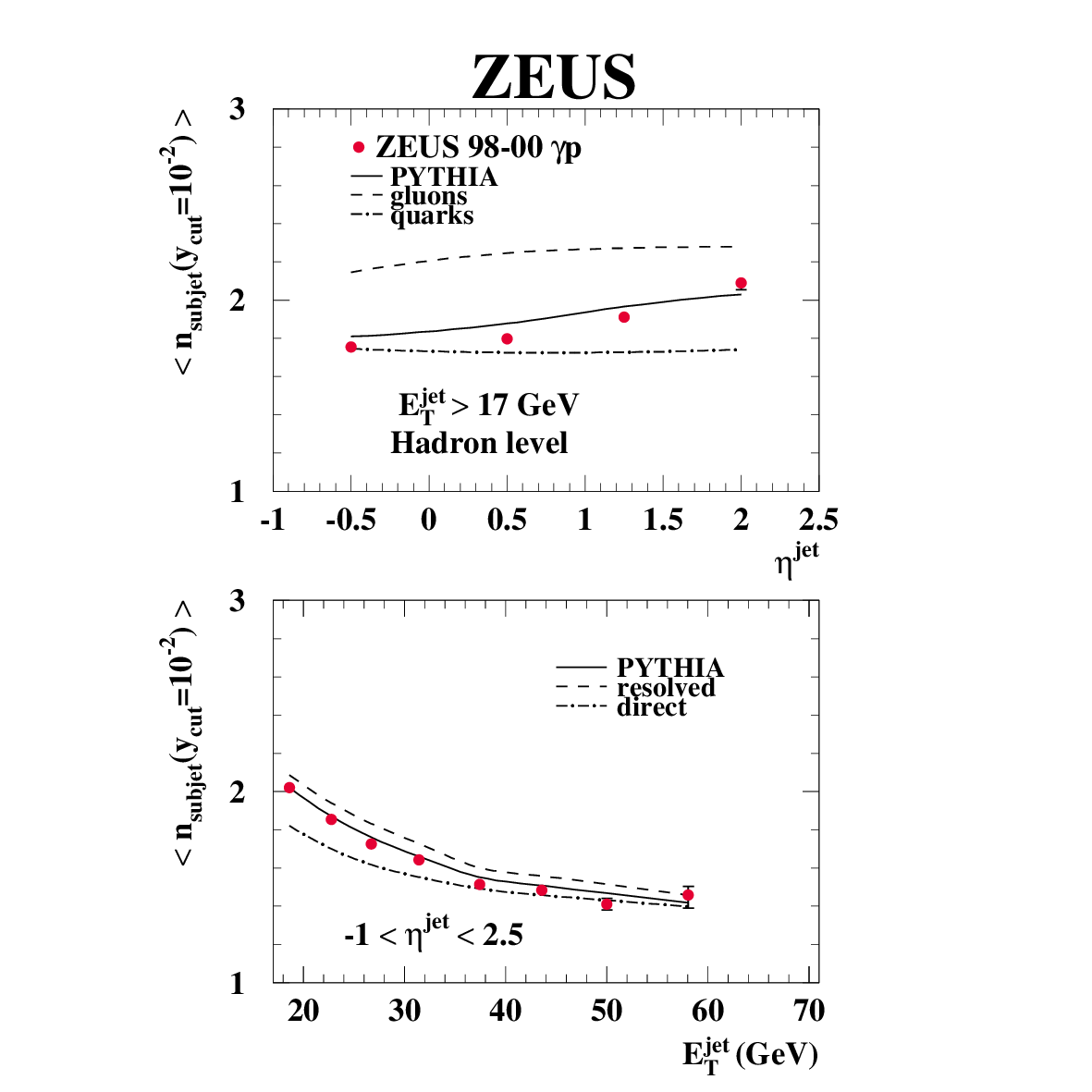,width=4.5cm}}
\put (4.0,2.5){\epsfig{figure=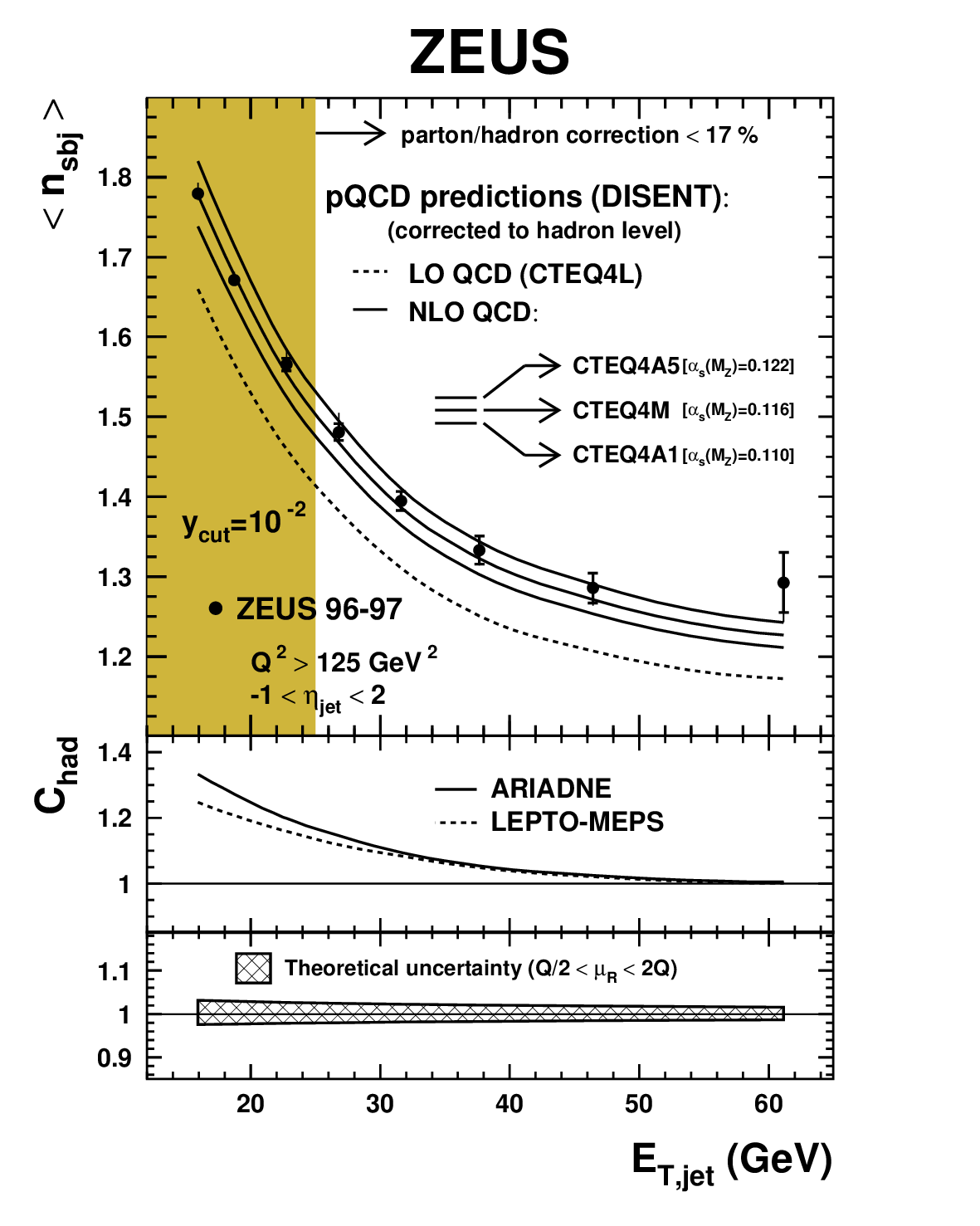,width=3.5cm}}
\put (2.0,-1.0){\epsfig{figure=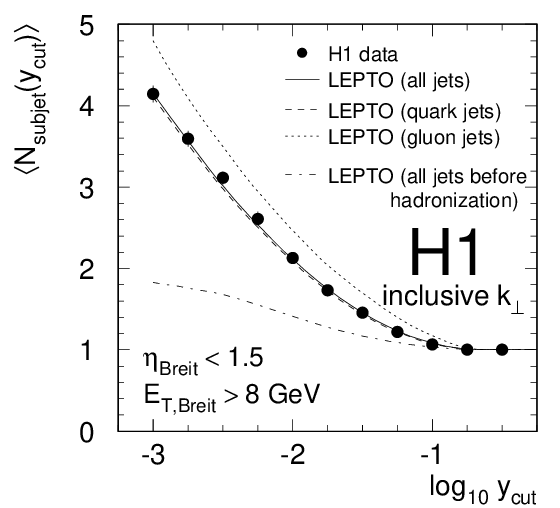,width=4cm}}
\end{picture}
\caption{\label{fig9}
{Mean subjet multiplicity in inclusive-jet photoproduction,
  inclusive-jet in NC DIS (laboratory frame) and dijet in NC DIS
  (Breit frame).}}
\end{figure}

The mean subjet multiplicity has been measured in inclusive-jet
photoproduction~\cite{np:b700:3}, dijets in NC DIS~\cite{np:b545:3}
and inclusive-jets in NC DIS~\cite{pl:b558:41} (see
Fig.~\ref{fig9}). The conclusions are similar to those from the
integrated jet shape. In addition, a value of $\asz$ was extracted
from this observable in NC DIS:
$\asmz{0.1187}{0.0017}{0.0009}{0.0024}{0.0076}{0.0093}$~\cite{pl:b558:41}.

%Figure 9

\section{Jet properties in different hard scattering processes}

\begin{figure}[h]
\setlength{\unitlength}{1.0cm}
\begin{picture} (18.0,2.0)
\put (0.5,-0.8){\epsfig{figure=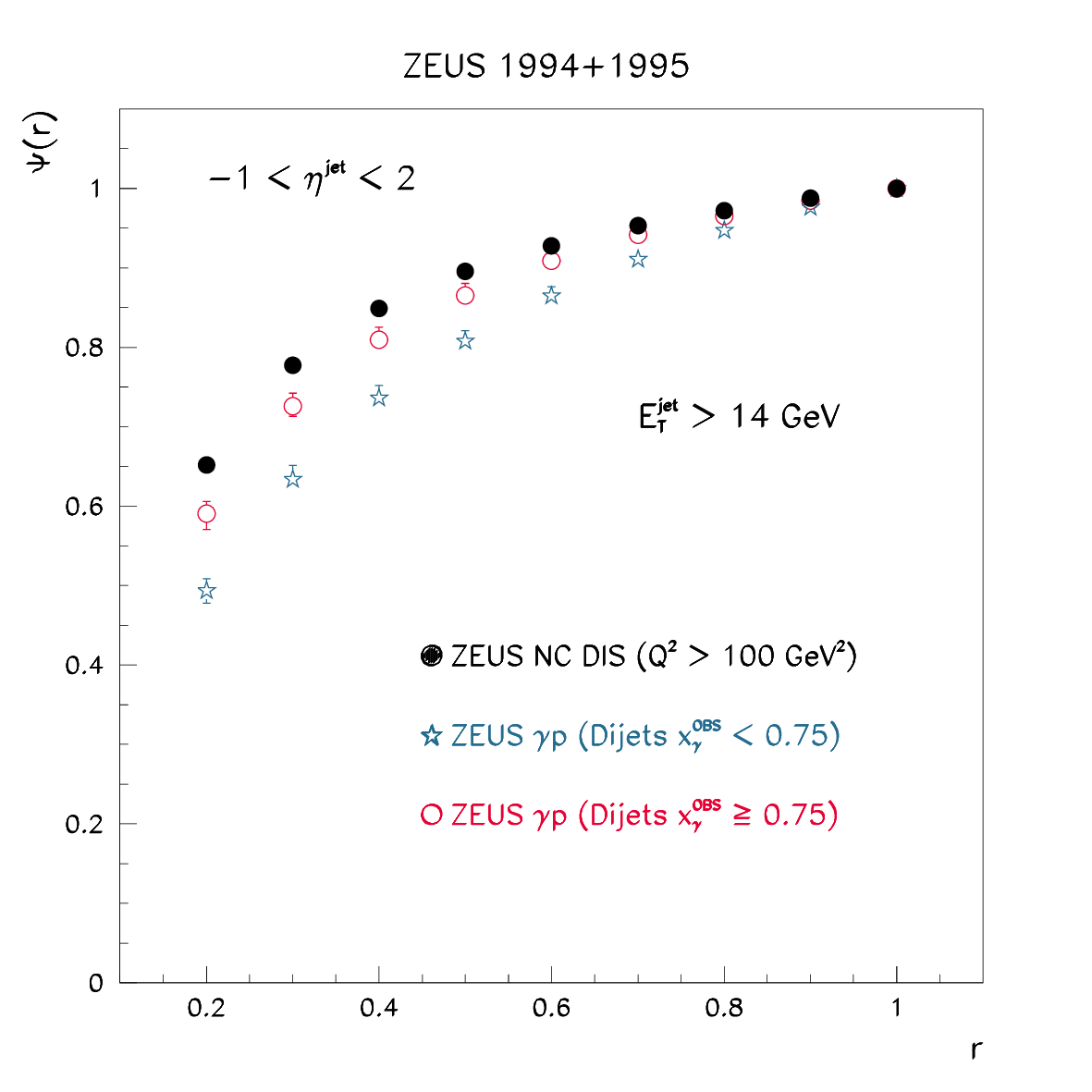,width=3.5cm}}
\put (3.5,-1.3){\epsfig{figure=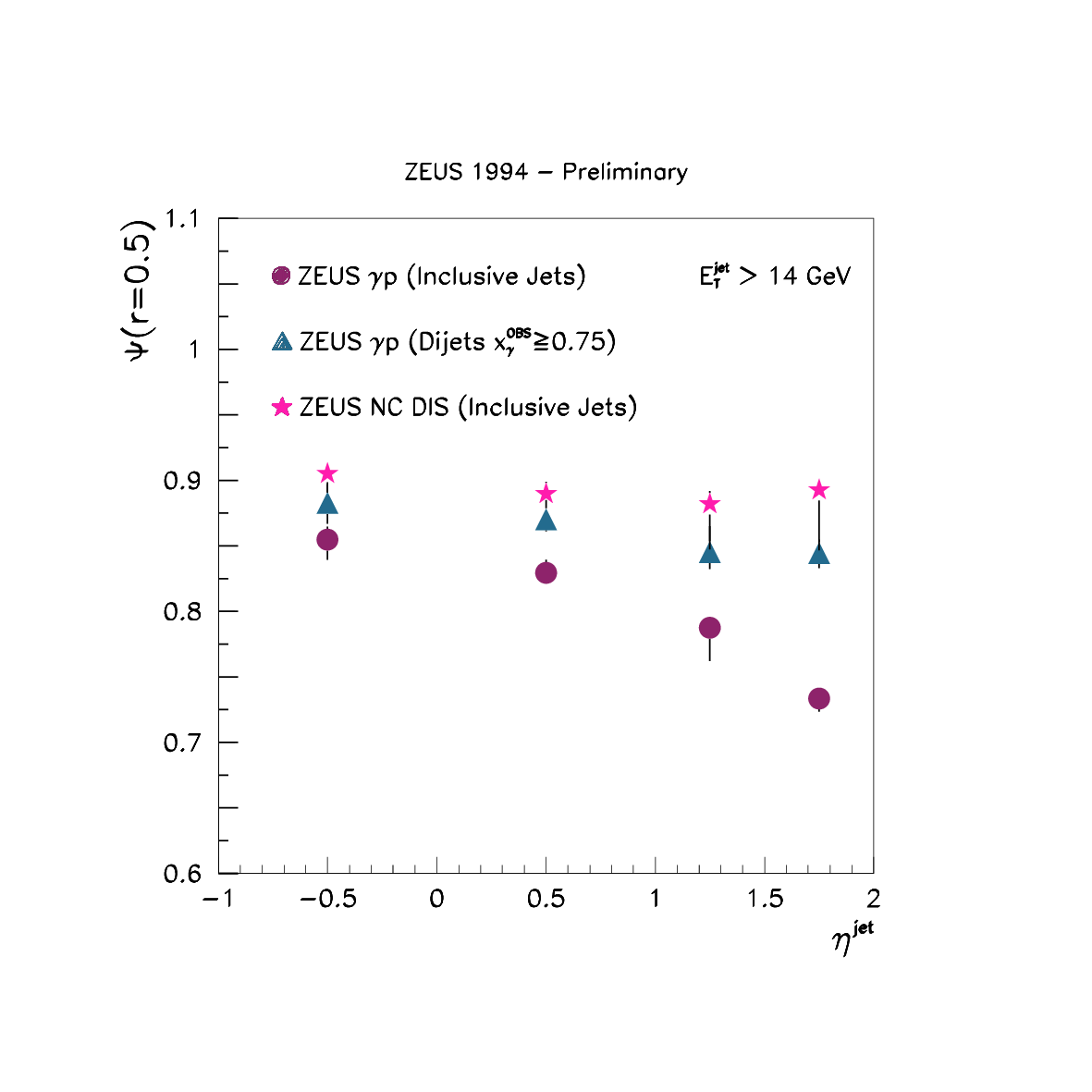,width=4.5cm}}
\put (5.66,2.4){\epsfig{figure=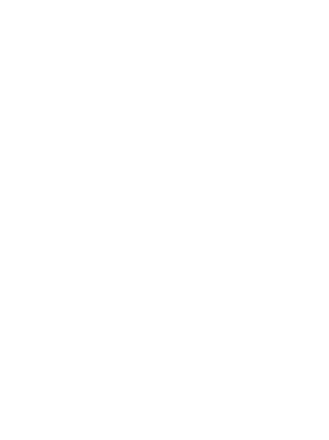,width=2.0cm,height=0.5cm}}
\put (2.0,-0.9){\tiny (a)}
\put (5.5,-0.9){\tiny (b)}
\end{picture}
\caption{\label{fig10}
{Mean integrated jet shape in inclusive-jet and dijet in
  photoproduction and inclusive-jet in NC DIS.}}
\end{figure}

The measurements of the integrated jet shape in photoproduction and NC
DIS were compared in order to study the properties of jets in
different hard scattering
processes~\cite{epj:c8:367}. Figure~\ref{fig10}a shows the comparison
of the jet shape in inclusive-jet NC DIS and dijets in photoproduction
with $\xo\lessgtr 0.75$ as a function of $r$. It is observed that the
jets in NC DIS, which are dominated by quark jets, are narrower than
those in resolved photoproduction, which include a larger fraction of
gluon jets, and similar to those in direct photoproduction, which are
also dominated by quark jets. Figure~\ref{fig10}b shows the integrated
jet shape for $r=0.5$ as a function of $\etajet$ for samples of
inclusive jets in photoproduction, dijets with $\xo>0.75$ in
photoproduction and inclusive jets in NC DIS. The inclusive-jet
photoproduction sample shows a behaviour which is consistent with an
increasing fraction of gluon jets as $\etajet$ increases, whereas the
dijets with $\xo>0.75$ and the NC DIS samples, which come
predominantly from quark jets, show no significant dependence with
$\etajet$.

%Figure 10

\begin{figure}[h]
\setlength{\unitlength}{1.0cm}
\begin{picture} (18.0,4.5)
\put (0.5,-1.0){\epsfig{figure=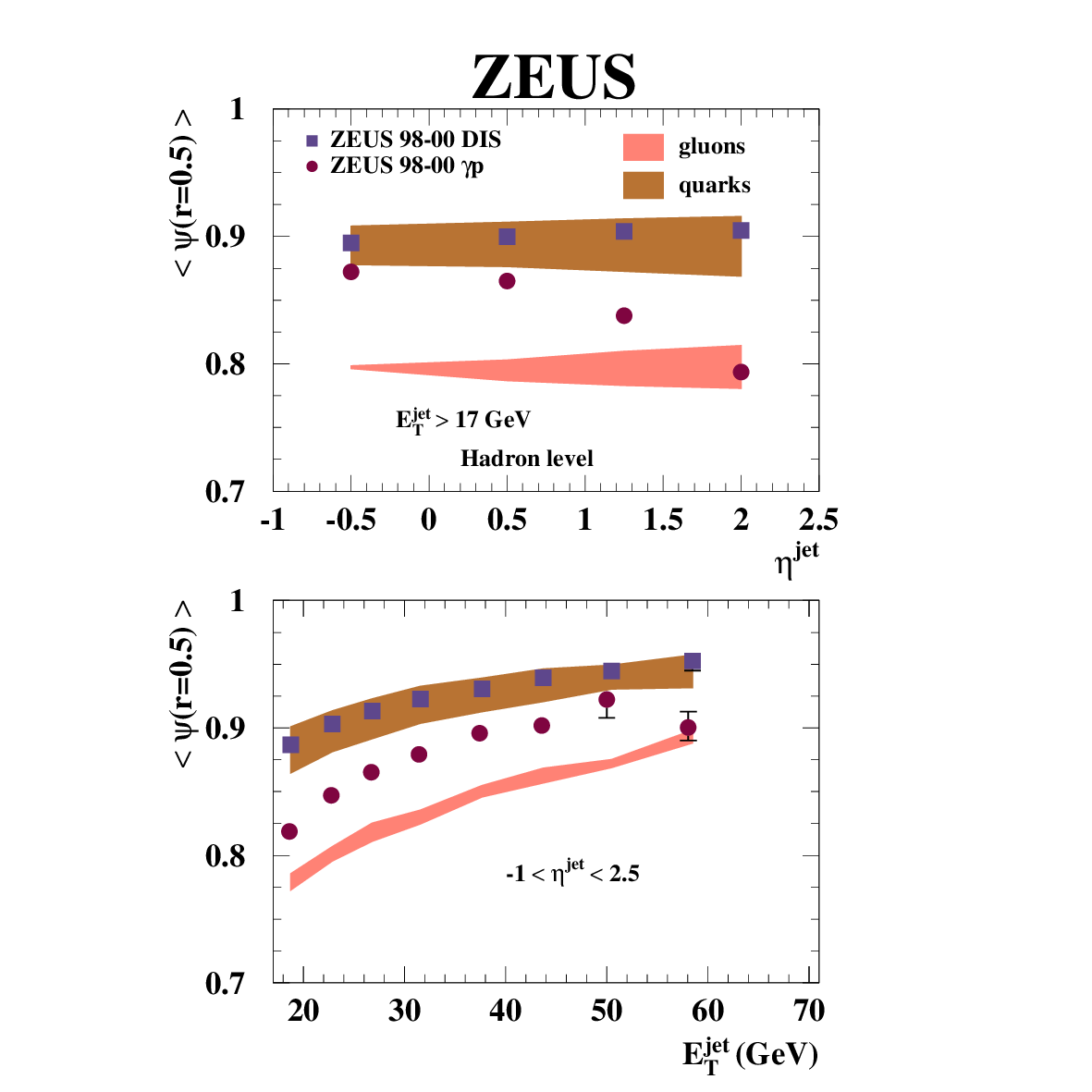,width=6cm}}
\end{picture}
\caption{\label{fig11}
{Mean integrated jet shape in inclusive-jet in photoproduction and NC
  DIS.}}
\end{figure}

The parton composition of the final state was studied in more detail
by comparing the $\etajet$ and $\etjet$ dependence of samples of
inclusive jets in photoproduction and NC DIS with the predictions of
pure quarks and gluons~\cite{np:b700:3} (see Fig.~\ref{fig11}). The NC
DIS sample is consistent with being dominated by quark jets in the
whole phase-space region. The jets in the photoproduction sample are
similar to those of NC DIS for low $\etajet$ and become increasingly
broader as $\etajet$ increases, which is consistent with the increase
of the fraction of gluon jets. For the $\etjet$ dependence, the NC DIS
jets are well described by the prediction of quark jets in the whole
measured range. The jets in the photoproduction sample get narrower
somewhat faster than those in NC DIS.

%Figure 11

\begin{figure}[h]
\setlength{\unitlength}{1.0cm}
\begin{picture} (18.0,4.0)
\put (1.0,-1.0){\epsfig{figure=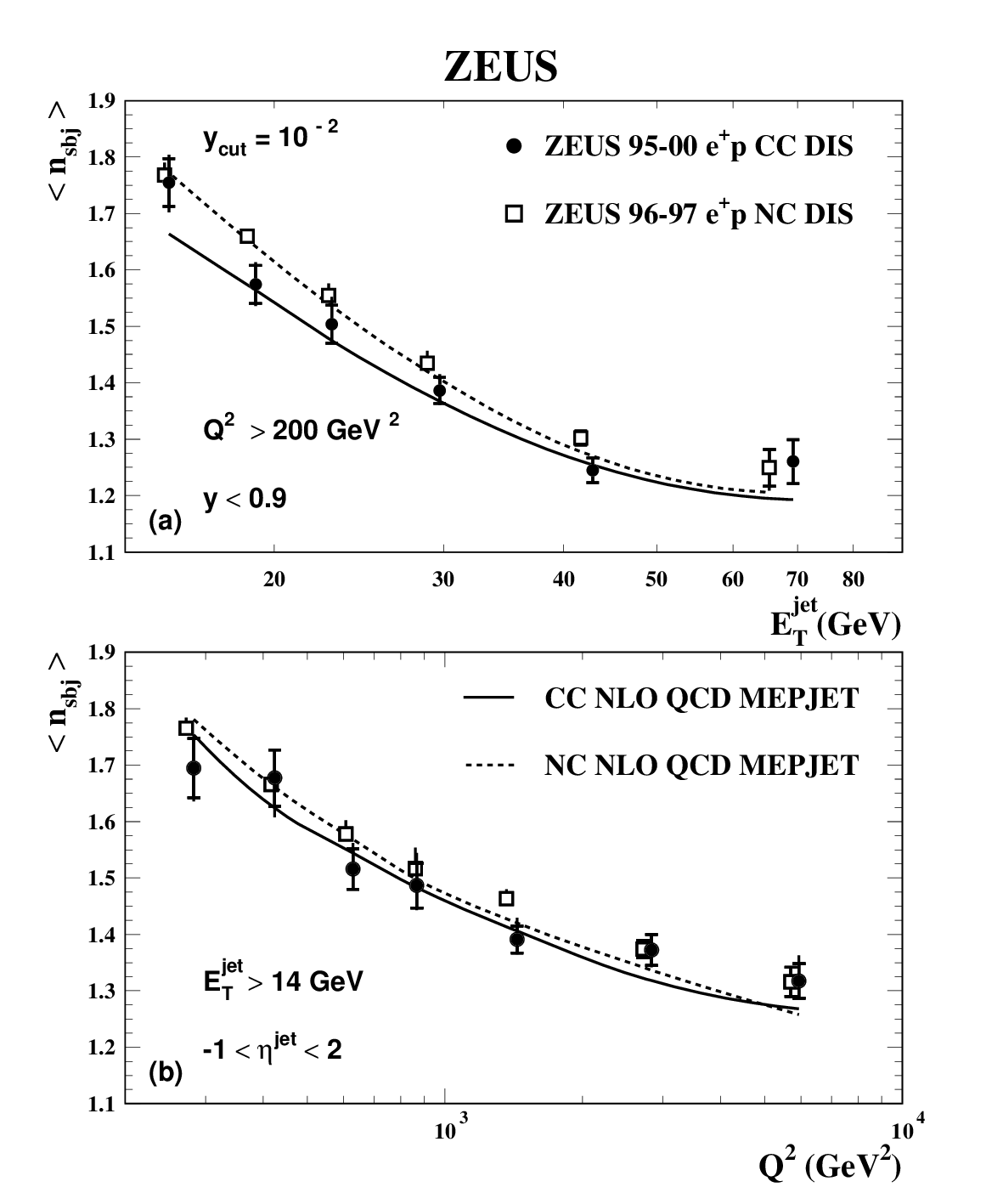,width=5cm}}
\end{picture}
\caption{\label{fig12}
{Mean subjet multiplicity in inclusive-jet in NC and CC DIS.}}
\end{figure}

The jets in the NC DIS sample were also compared to those of an
inclusive-jet sample in CC DIS~\cite{epj:c31:149}, also dominated by
quark jets, so a similar dependence with e.g. $\etjet$ is expected. 
Figure~\ref{fig12} shows the mean subjet multiplicity for a fixed
value of $\yc$ as functions of $\etjet$ and $\q2$ in NC and CC
DIS. The jets in both samples get narrower as $\etjet$ increases, but
the jets in CC seem to be slightly narrower than those in NC, in
agreement with the NLO predictions~\cite{pl:b380:205}. These
differences can be understood in terms of the different $\q2$ spectra
in NC and CC DIS.

%Figure 12

\begin{figure}[h]
\setlength{\unitlength}{1.0cm}
\begin{picture} (18.0,3.5)
\put (1.0,-1.0){\epsfig{figure=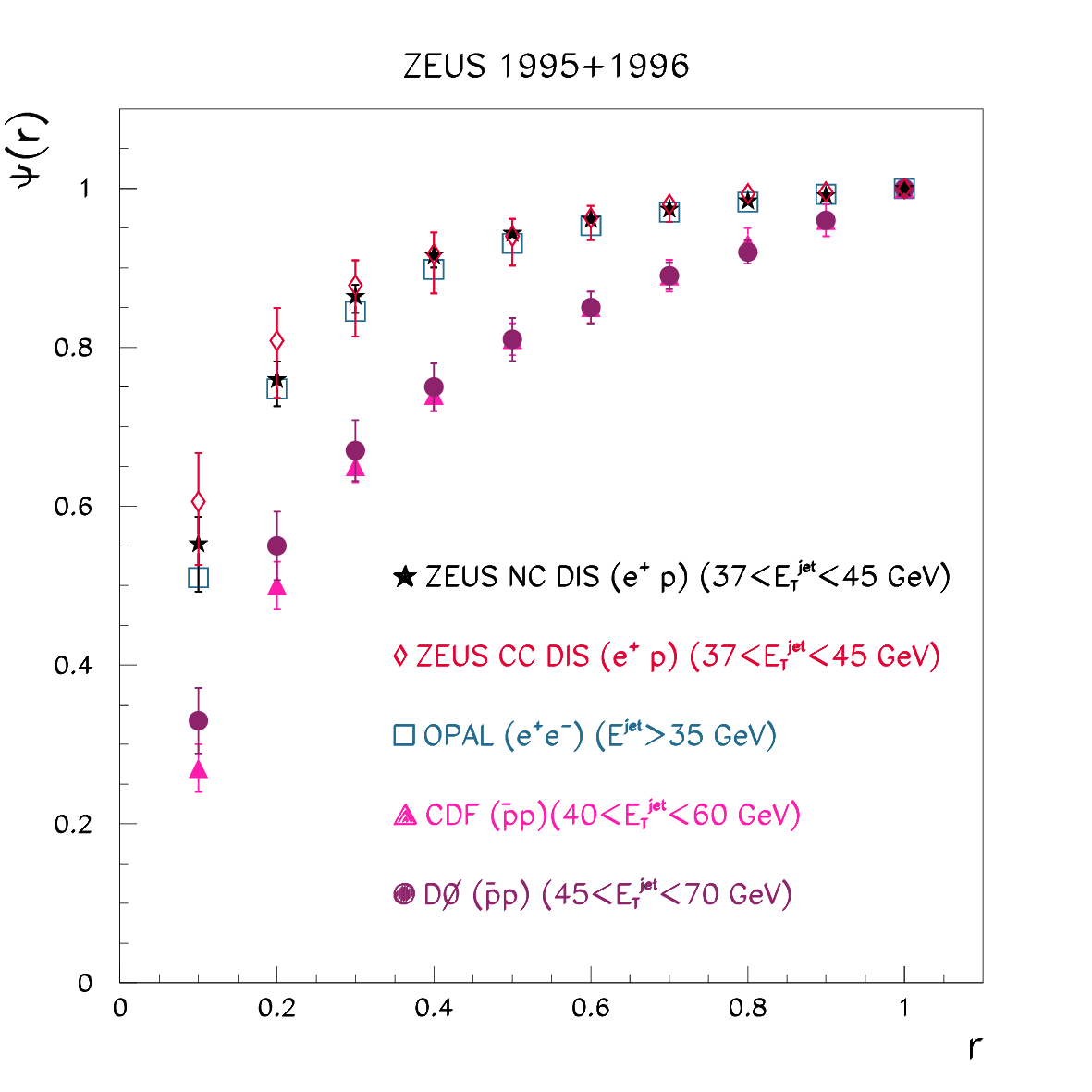,width=5cm}}
\end{picture}
\caption{\label{fig13}
{Mean integrated jet shape in inclusive-jet in $\ele$, $\pp$ and NC
  and CC DIS.}}
\end{figure}

The measurements in inclusive jets in NC DIS~\cite{epj:c8:367} were
compared to similar measurements from CDF~\cite{prl:70:713},
D\O~\cite{pl:b357:500} and OPAL~\cite{zfp:c63:197}. The jets in NC
DIS and $\ele$ are similar (see Fig.~\ref{fig13}), which is understood
in terms of a similar parton composition of the final state, dominated
by quark jets. The jets in NC DIS and $\ele$ are narrower than
those in $\pp$ collisions, which can be understood from the fact that
the final-state jets in $\pp$ come predominantly from gluons, as in
resolved photoproduction. From these comparisons and the ones shown
before, it can be concluded that the pattern of QCD radiation within a
quark jet is to a large extent independent of the hard scattering
process.

%Figure 13

\section{Quark- and gluon-jet properties}

Several analyses were done at HERA in which quark and gluon jets were
isolated and the dynamics of the underlying subprocesses
studied. In some of these analyses, quark and gluon jets were
identified by using samples of dijet events in photoproduction which
contain charm quarks in the final state. In these analyses, one of the
jets is tagged as a charm and the substructure of the other
(``untagged'') jet was studied. This method provides an unbiased
sample of quark jets. In direct photoproduction, the untagged jet is
also a charm quark, but in resolved processes, since there are several
contributing processes, such as gluon-gluon fusion or charm-excitation
processes, the untagged jet can be a quark or a gluon.

\begin{figure}[h]
\setlength{\unitlength}{1.0cm}
\begin{picture} (18.0,2.0)
\put (0.0,-0.8){\epsfig{figure=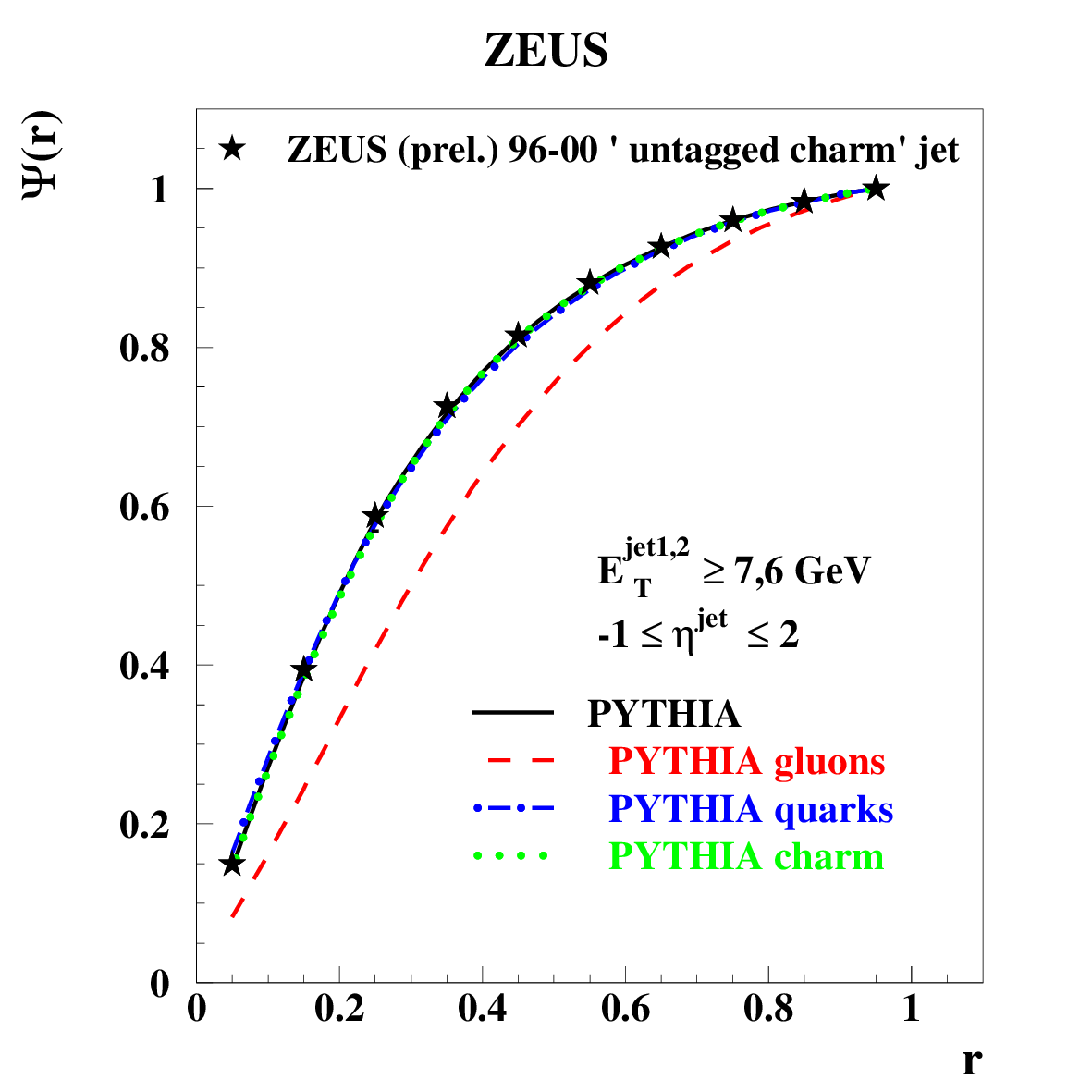,width=3.5cm}}
\put (3.5,-0.8){\epsfig{figure=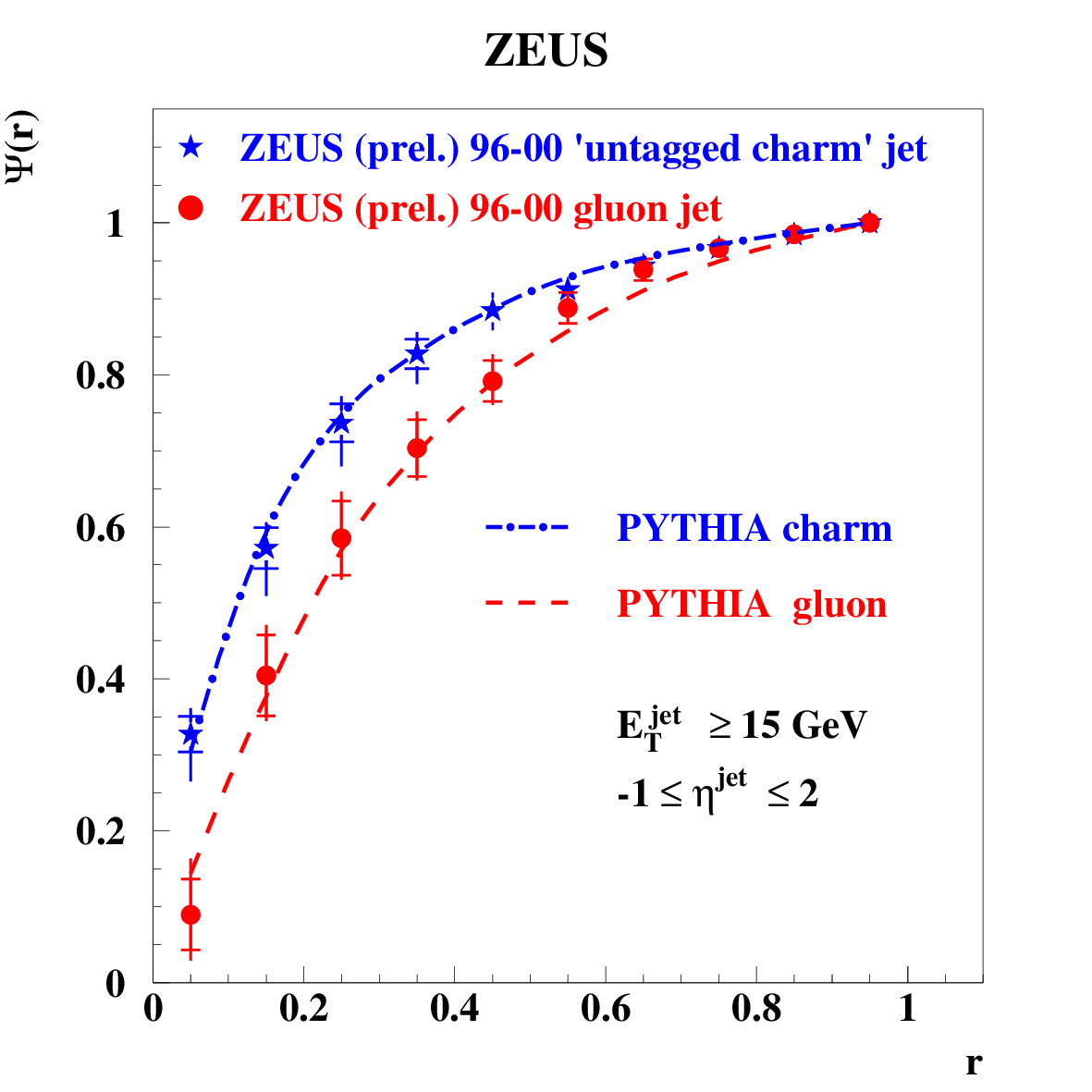,width=3.5cm}}
\put (1.5,-0.9){\tiny (a)}
\put (5.5,-0.9){\tiny (b)}
\end{picture}
\caption{\label{fig14}
{Mean integrated jet shape in charm photoproduction.}}
\end{figure}

The ZEUS Collaboration identified quark and gluon jets by selecting a
sample of dijet events in photoproduction and tagging the subsample
with charm production via the identification of a $D^*$ meson 
associated with one of the jets~\cite{zeus:prelim:2001}. The
substructure of the untagged jet was measured and found to be
consistent with the predictions for quark-initiated jets (see
Fig.~\ref{fig14}a). The substructure of gluon jets was extracted by
means of 
${\cal O}_{\rm dijet}=f_q\cdot{\cal O}_{\rm quark}+f_g \cdot{\cal O}_{\rm gluon}$,
where the substructure of dijets is the measured observable,
the substructure of quarks was approximated using the measured
substructure of untagged charm jets and the fraction of quarks and
gluons was estimated using the MC predictions, which describe the data
well. Figure~\ref{fig14}b shows the extracted substructure of gluon
jets compared with the measured substructure of untagged jets. The
gluon jets are observed to be broader than the quark jets. The model
predictions describe the measurements well.

%Figure 14

\begin{figure}[h]
\setlength{\unitlength}{1.0cm}
\begin{picture} (18.0,2.8)
\put (0.5,3.0){\epsfig{figure=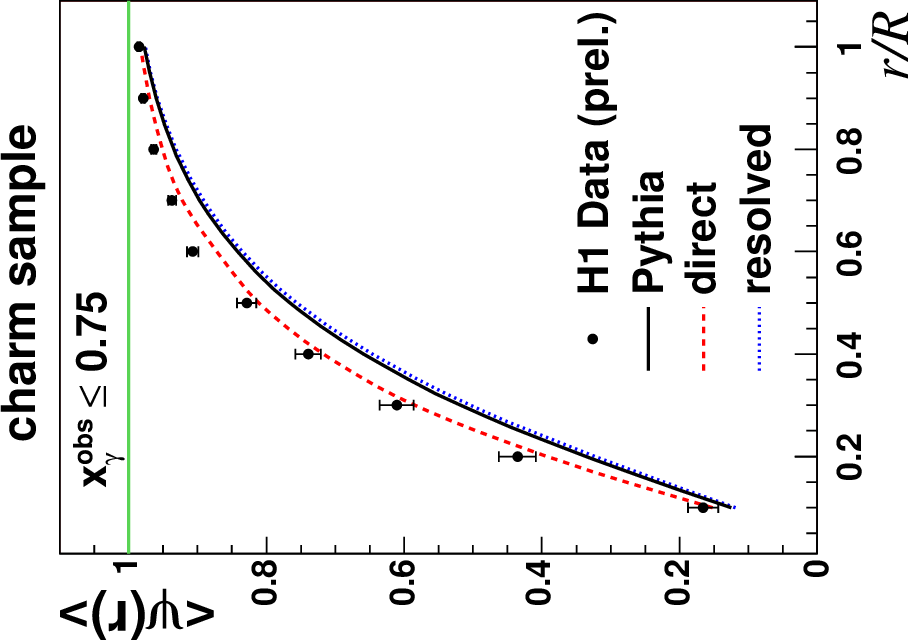,width=4cm,angle=-90}}
\put (3.5,3.0){\epsfig{figure=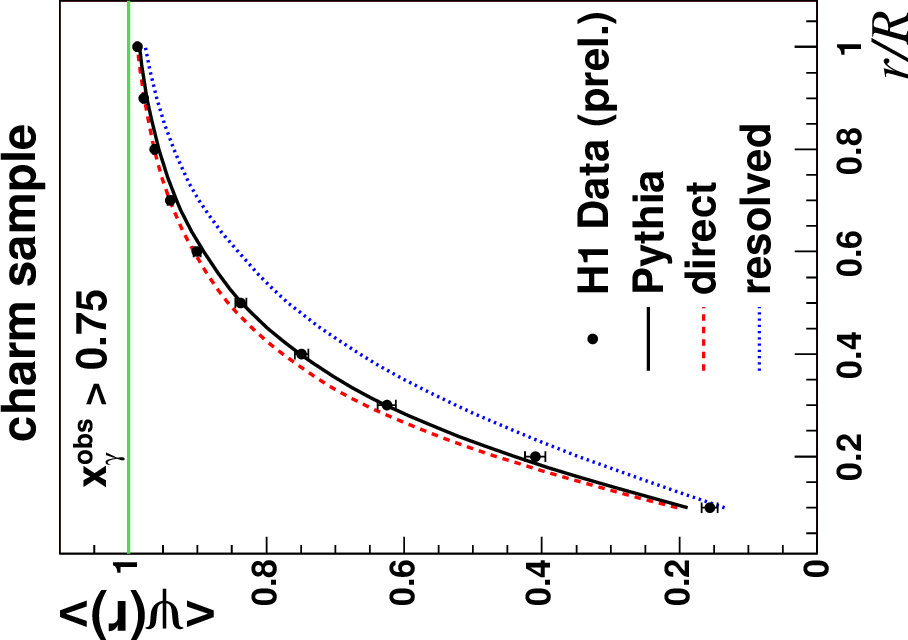,width=4cm,angle=-90}}
\end{picture}
\caption{\label{fig15}
{Mean integrated jet shape in charm photoproduction.}}
\end{figure}

The H1 Collaboration made a similar analysis, but tagging the charm
jets by associating a muon to one of the
jets~\cite{h1-prelim-05-077}. The substructure of the other jet was
studied for $\xo\lessgtr 0.75$ (see Fig.~\ref{fig15}). The QCD-based
MC predictions describe the data well for $\xo>0.75$. However, some
differences are observed for $\xo<0.75$. The data suggest a smaller
fraction of gluon jets, which can come only from the gluons in the
charm-excitation subprocess, at low $\xo$ than the predictions.

%Figure 15

\begin{figure}[h]
\setlength{\unitlength}{1.0cm}
\begin{picture} (18.0,2.0)
\put (-0.5,-1.5){\epsfig{figure=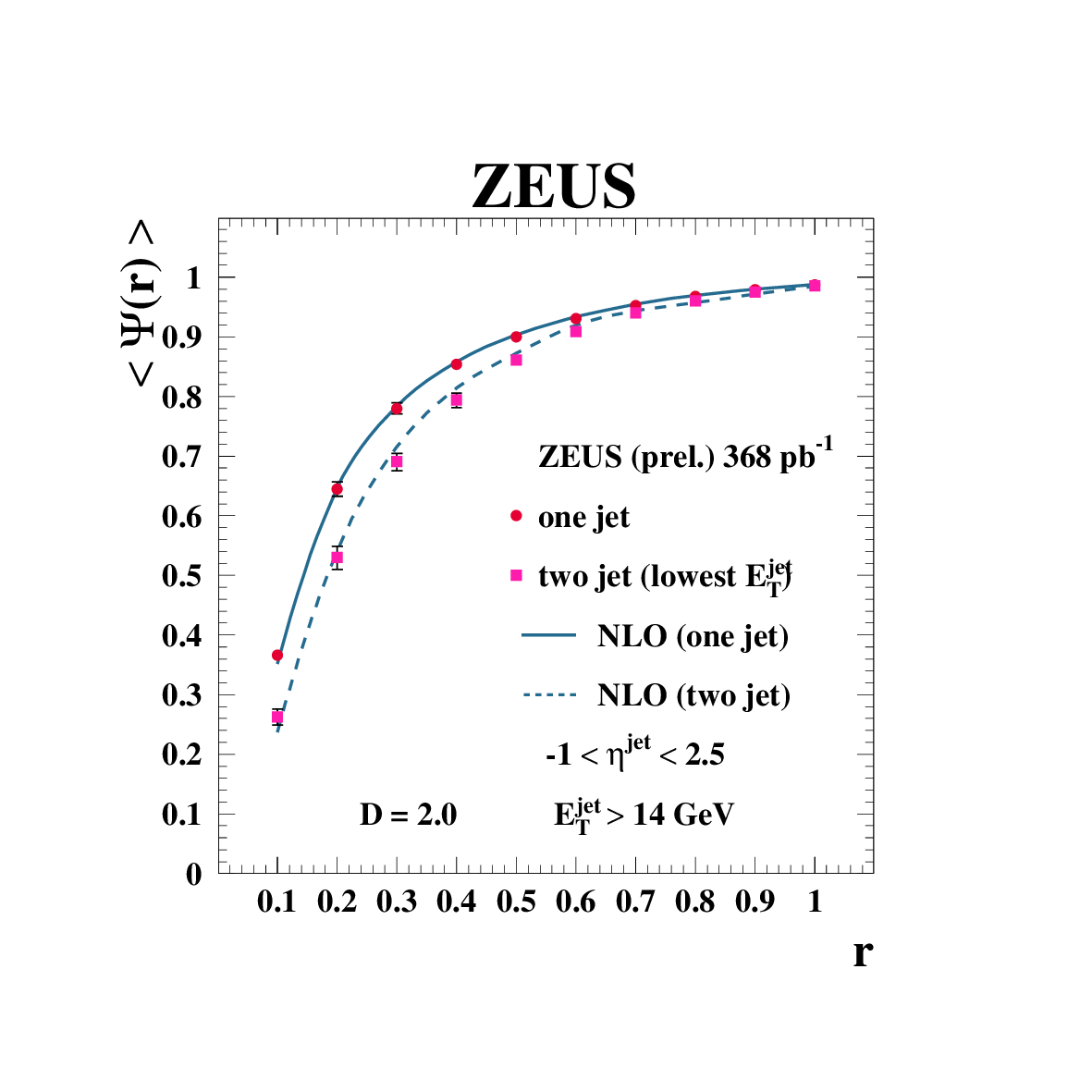,width=5cm}}
\put (3.5,-1.5){\epsfig{figure=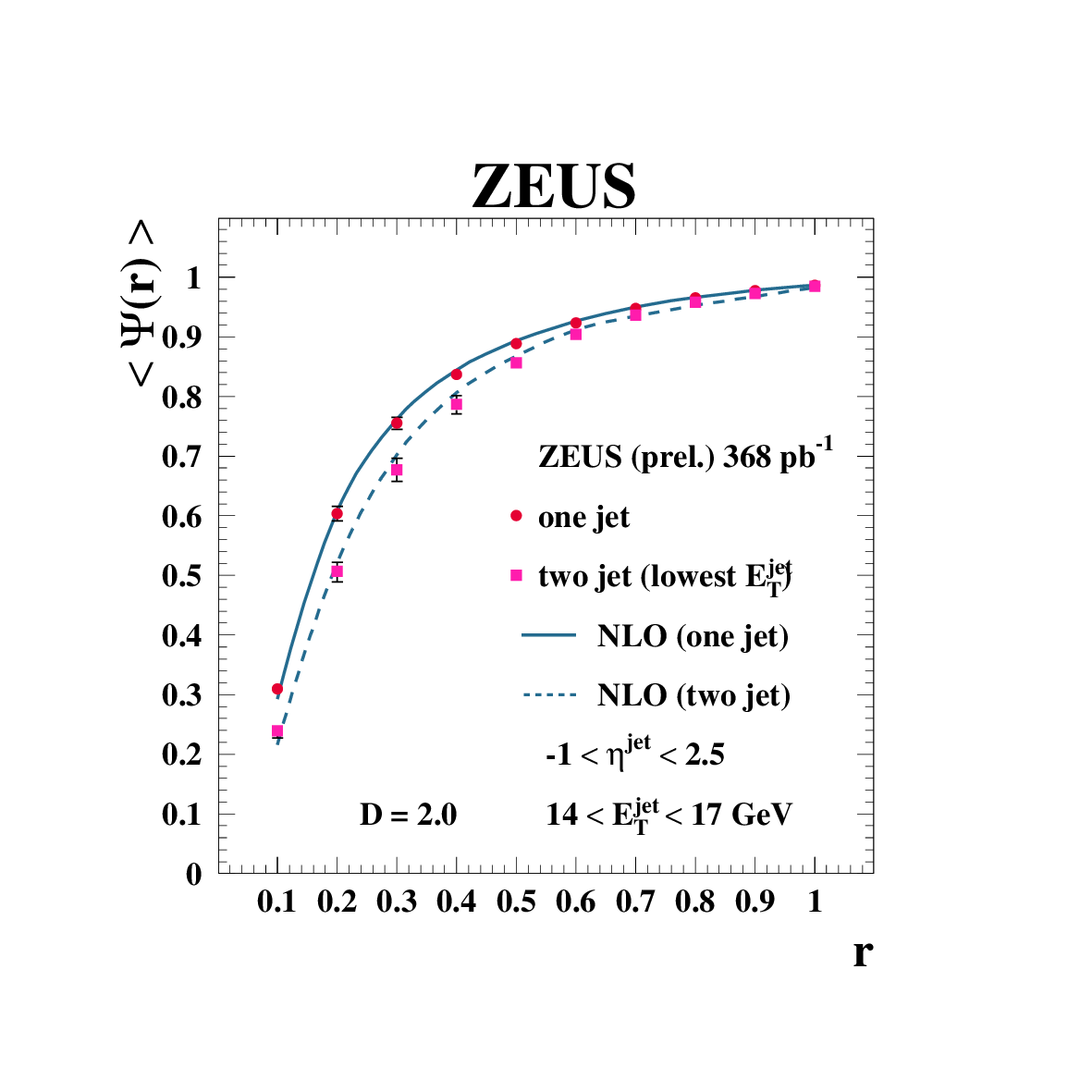,width=5cm}}
\end{picture}
\caption{\label{fig16}
{Mean integrated jet shape in one-jet and two-jet in NC DIS.}}
\end{figure}

The differences between quark and gluon jets were also studied by
exploiting the different type of parton content in the final state for
one-jet and two-jet samples in NC DIS in the laboratory
frame~\cite{zeus-prel-07-013}. The one-jet sample is expected to be
dominated by quark jets, whereas the two-jet sample is expected to
contain a larger fraction of gluon jets. Figure~\ref{fig16} shows the
measurement of the mean integrated jet shape as a function of $r$ for
$\etjet>14$~GeV and for $14<\etjet<17$~GeV for the one-jet and the
two-jet samples. For the two-jet sample, the substructure of the
lowest-$\etjet$ jet was studied for those events in which the two jets
are closer than two units in the $\eta-\phi$ plane. These jets are
observed to be broader than the jets in the one-jet sample. The data
are compared to NLO calculations~\cite{np:b485:291,prl:87:082001}. The
calculations give a good description of the data and show that the
measurements are consistent with a higher gluon content in the two-jet
sample.

%Figure 16

\begin{figure}[h]
\setlength{\unitlength}{1.0cm}
\begin{picture} (18.0,3.0)
\put (0.5,-1.5){\epsfig{figure=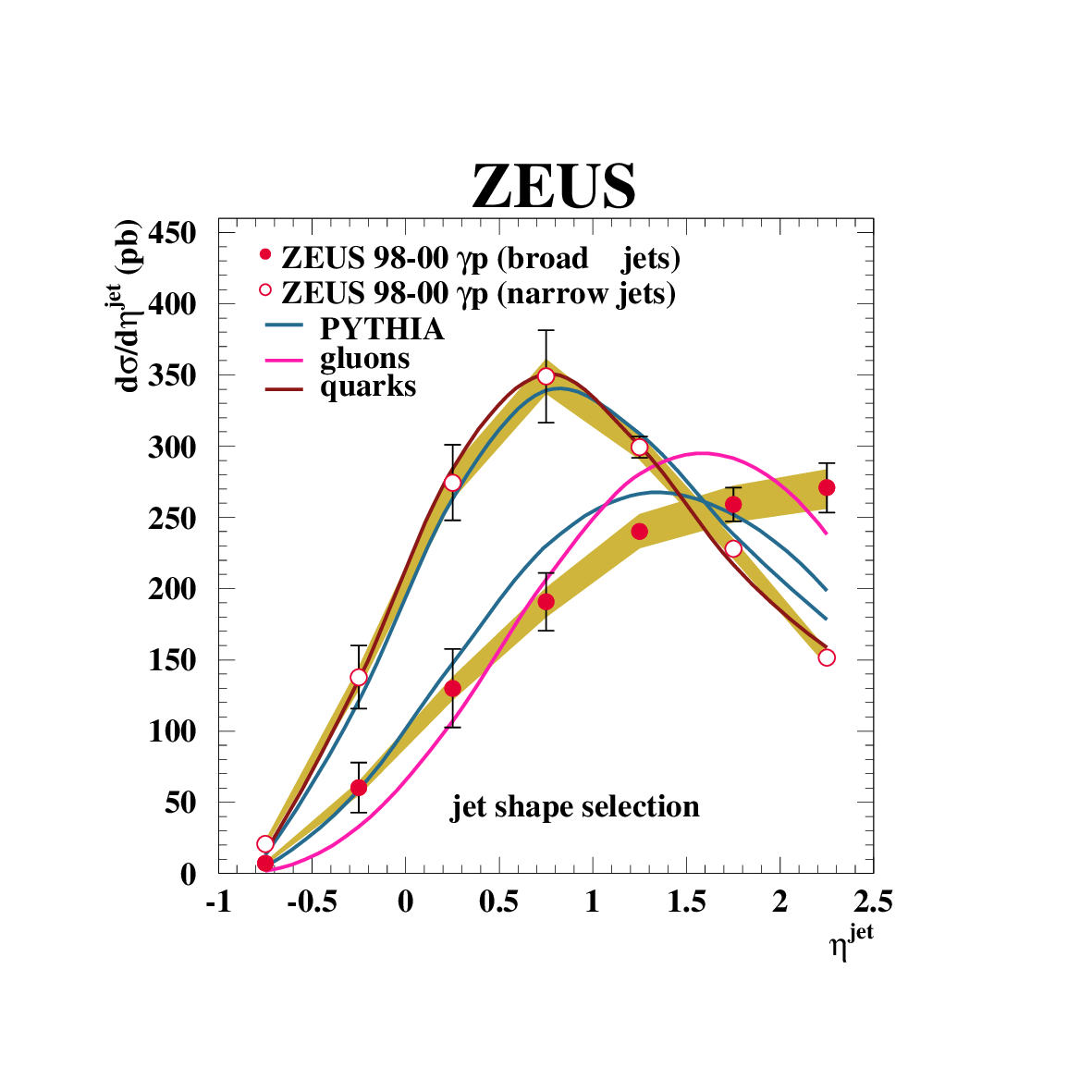,width=6cm}}
\end{picture}
\caption{\label{fig17}
{Inclusive-jet cross sections in photoproduction.}}
\end{figure}

Since the QCD prediction that gluon jets are broader than quark
jets has been amply proven by data, it is possible to exploit this
fact for identifying the different type of jets and study the dynamics
of the subprocesses. Quark and gluon jets were
identified~\cite{np:b700:3} on a statistical basis by selecting
``narrow'' and ``broad'' jets by requiring that the measured jet shape
is smaller than 0.6 or larger than 0.8, respectively. Using this
selection, the inclusive-jet cross section in photoproduction was
measured as a function of $\etajet$ (see Fig.~\ref{fig17}). The
measured cross section for broad jets displays a very different shape
than that of the narrow jet sample. The $\etajet$ distribution for the
narrow jet sample peaks at about 0.5, whereas the broad jet sample
distribution increases as $\etajet$ increases. The predictions from
the QCD-based MC models, with the same jet shape selection, give a
good description of the shape of the data. The MC predicts that the
broad jet sample is dominated by $qg\rightarrow qg$ via gluon exchange
in resolved processes, whereas the narrow jet sample is dominated by
boson-gluon fusion events in direct processes. Therefore, the narrow
jet sample is indeed dominated by quark-initiated jets in the final
state and the broad jet sample has a higher content of gluon-initiated
jets. The comparison with the predictions for samples of pure quark or
gluon jets supports this conclusion.

%Figure 17

\begin{figure}[h]
\setlength{\unitlength}{1.0cm}
\begin{picture} (18.0,2.0)
\put (0.0,-1.3){\epsfig{figure=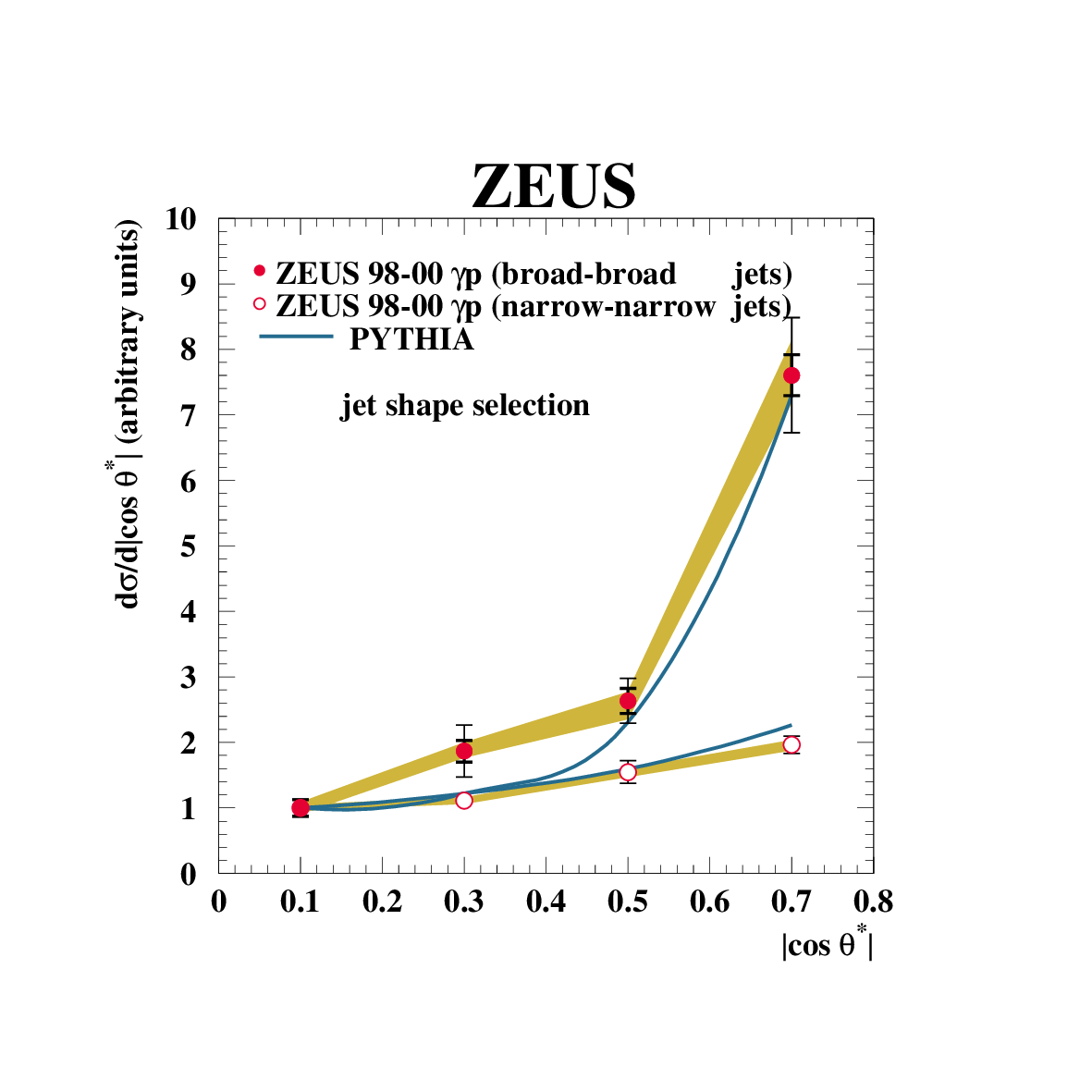,width=4.5cm}}
\put (3.5,-1.3){\epsfig{figure=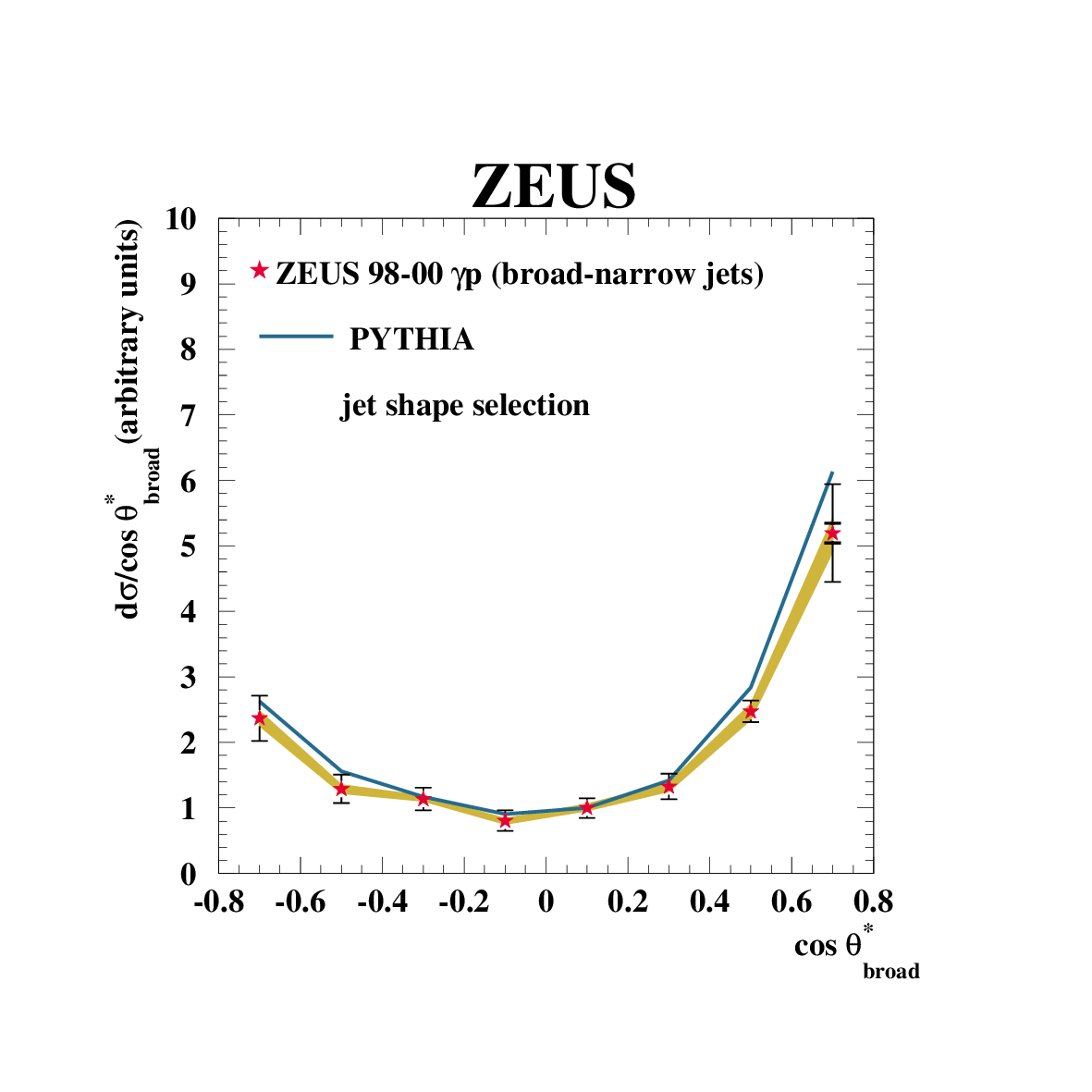,width=4.5cm}}
\put (1.9,-0.9){\tiny (a)}
\put (5.5,-0.9){\tiny (b)}
\end{picture}
\caption{\label{fig18}
{Dijet cross sections in photoproduction.}}
\end{figure}

The underlying parton dynamics is reflected in the distribution of
the scattering angle in the dijet centre-of-mass system,
$\theta^*$. The $\cos(\theta^*)$ distribution is sensitive to the spin
of the exchanged particle: a different behaviour is expected for quark
and gluon exchange. Samples selected according to the internal
structure of the jets give an unbiased handle to explore the
dynamics of the subprocesses. Figure~\ref{fig18}a shows the dijet
cross section as a function of $\cos(\theta^*)$ for samples of two
broad or two narrow jets. The cross section for two broad jets rises
more steeply than the sample of two narrow jets. The MC predictions
give a reasonable description of the data. The different slope
observed in the data can be understood in terms of the dominant
subprocesses in the two samples: the two broad jet sample is dominated
by subprocesses mediated by gluon exchange whereas the two narrow jet
sample is dominated by subprocesses mediated by quark exchange. The
$\cos(\theta^*)$ distribution for a sample of one narrow and one 
broad jet in the same event shows a clear asymmetry (see
Fig.~\ref{fig18}b). The MC reproduces the data well. The asymmetry
observed can be understood in terms of the dominant subprocess: 
$qg\rightarrow qg$, the positive side is dominated by $t$-channel
gluon exchange, whereas the negative side is dominated by $u$-channel
quark exchange.

%Figure 18

\section{The pattern of parton radiation}

\begin{figure}[h]
\setlength{\unitlength}{1.0cm}
\begin{picture} (18.0,3.6)
\put (1.0,-1.3){\epsfig{figure=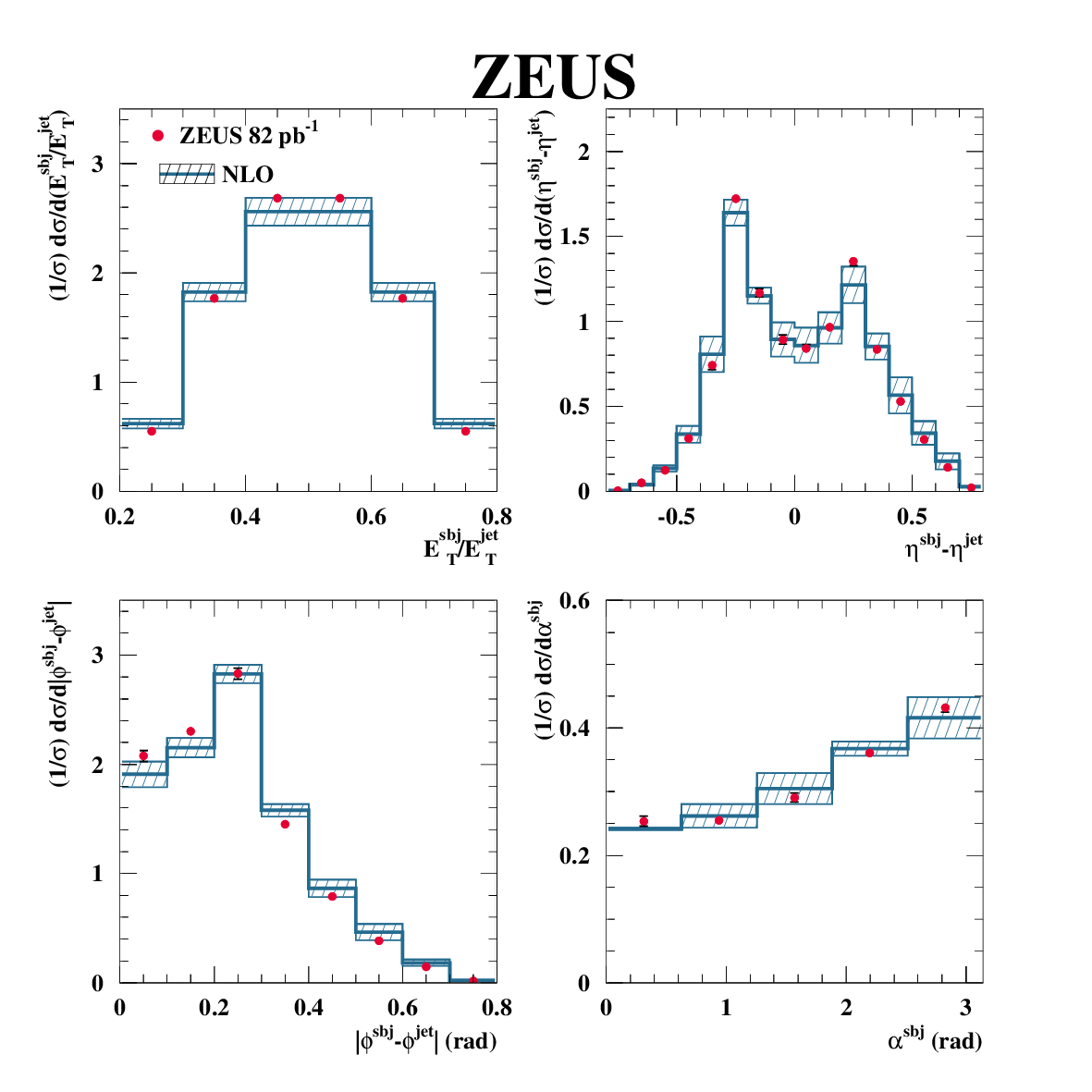,width=6cm}}
\end{picture}
\caption{\label{fig19}
{Normalised subjet cross sections in inclusive-jet NC DIS.}}
\end{figure}

The pattern of parton radiation within a jet was studied using subjet
distributions. The pattern of parton radiation from a primary parton
is dictated in QCD by the splitting functions, so the measurements
provide a direct test of these functions and their scale
dependence. The subjet distributions can also be used to study
colour-coherence effects between the initial and final states. QCD
predicts that soft gluon radiation tends to be emitted towards the
proton direction. The measurements of subjet distributions were
done in NC DIS~\cite{zeus-pub-08-014} as functions of the fraction of
subjet transverse energy, $\etsbj/\etjet$, the difference between the
$\eta$ and $\phi$ of the subjet with respect to the jet,
$\etasbj\!-\!\etajet$ and $|\phisbj\!-\!\phijet|$, and $\asbj$, the
angle, as viewed from the jet centre, between the subjet with
higher-$E_T$ and the proton beam line in the $\eta-\phi$ plane. The
cross sections were measured for those jets which contain exactly two
subjets for $\yc=0.05$. Figure~\ref{fig19} shows the normalised subjet
cross sections. The measurements show that the two subjets tend to
have similar transverse energy. The distribution in
$\etasbj\!-\!\etajet$ has a two-peak structure. The cross section
as a function of $|\phisbj\!-\!\phijet|$ shows a suppression around
$\Delta\phi=0$, which comes from the fact that the two subjets
cannot be resolved when they are too close together; this is also
observed in the dip between the two peaks in
$\etasbj\!-\!\etajet$. The $\asbj$ distribution shows that the
highest-$E_T$ subjet tends to be in the rear direction. This is
consistent with the asymmetric two peak structure observed in the
$\etasbj\!-\!\etajet$ distribution and related to colour-coherence
effects. NLO QCD predictions~\cite{np:b485:291} are compared to the
data and give an adequate description.

%Figure 19

\begin{figure}[h]
\setlength{\unitlength}{1.0cm}
\begin{picture} (18.0,2.5)
\put (1.0,-1.8){\epsfig{figure=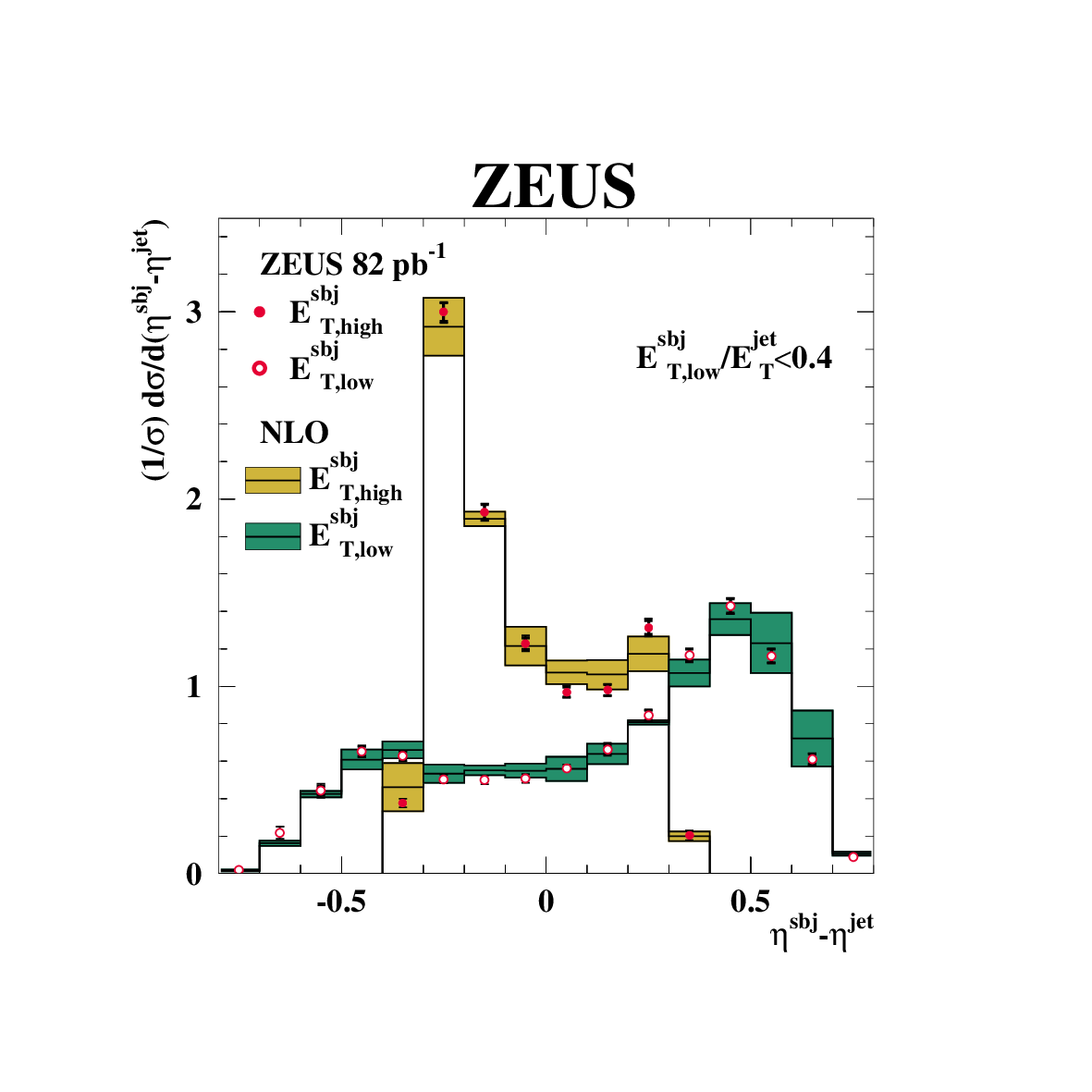,width=6cm}}
\end{picture}
\caption{\label{fig20}
{Normalised subjet cross sections in inclusive-jet NC DIS.}}
\end{figure}

To study in more detail the colour-coherence effects, the cross
section as a function $\etasbj\!-\!\etajet$ was measured for those
jets with subjets of different transverse energy fraction, separately
for the low- and high-$E_T$ subjets. The measurements show that the
high-$E_T$ subjet tends to be in the rear whereas the low-$E_T$ 
subjet tends to be emitted in the forward direction (see
Fig.~\ref{fig20}). The NLO predictions describe the data well. This
behaviour can be attributed to colour-coherence effects between the
initial and final states, and indicates that soft gluon radiation is
emitted predominantly towards the proton direction.

%Figure 20

\begin{figure}[h]
\setlength{\unitlength}{1.0cm}
\begin{picture} (18.0,4.3)
\put (1.0,-1.3){\epsfig{figure=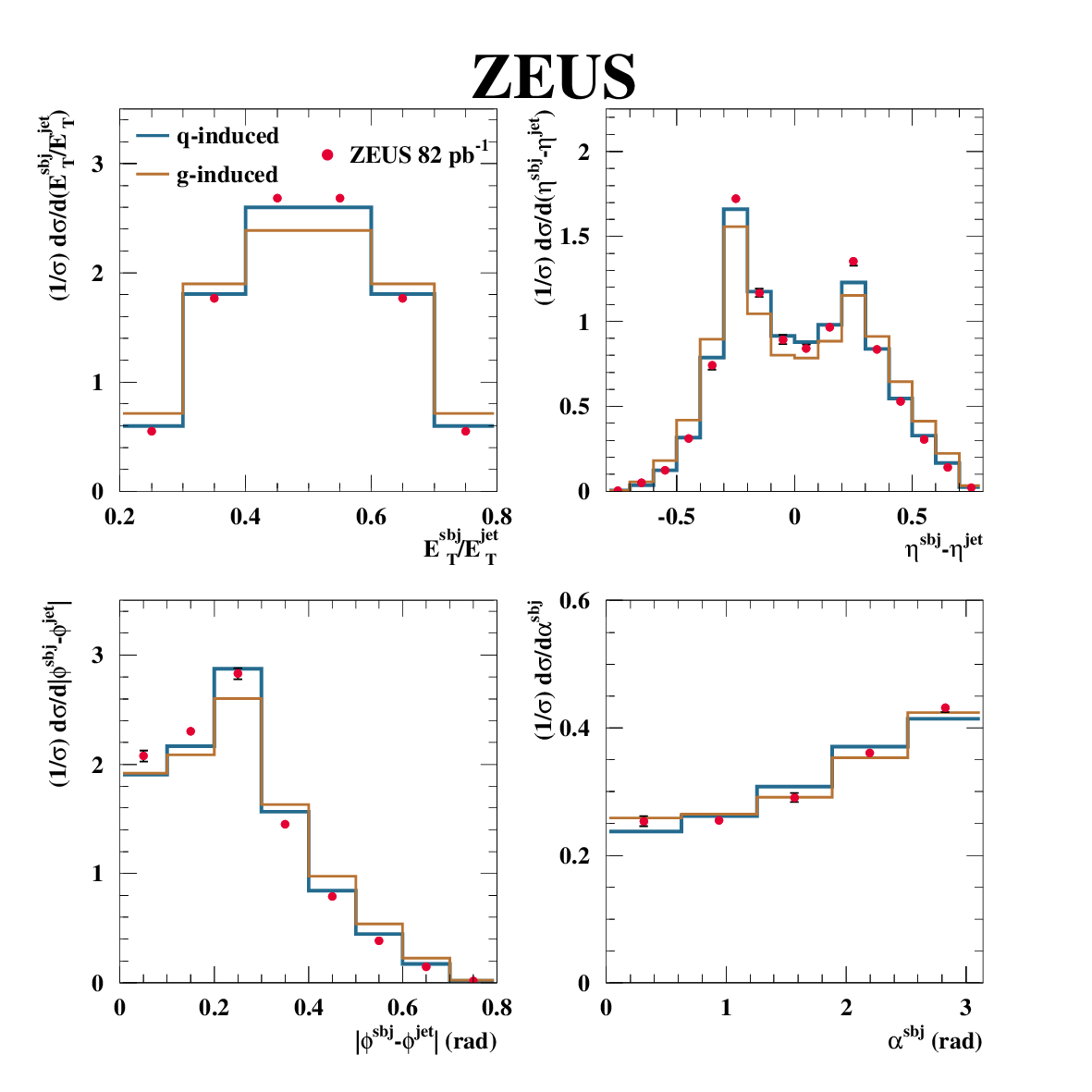,width=6cm}}
\end{picture}
\caption{\label{fig21}
{Normalised subjet cross sections in inclusive-jet NC DIS.}}
\end{figure}

The pattern of parton radiation was further studied by comparing
the predictions for quark- and gluon-induced splittings separately
with the data. The NLO calculations predict that the rate is dominated
by quark-induced processes. The predictions of these two types of
processes have a different shape: the quark-induced processes have
subjets with more similar transverse energy than the subjets from the
gluon-induced processes, and the two subjets coming from a $qg$ pair
are closer together than in the case of $\qq$ pairs. The comparison
with the measurements shows that the data are better described by the 
calculations for jets arising from the splitting of a quark into a $qg$
pair (see Fig.~\ref{fig21}).

%Figure 21

\section{Summary}
Jet substructure has been extensively studied at HERA in terms of jet
shapes and subjet multiplicities and distributions. The measurements
performed allowed stringent tests of perturbative QCD directly beyond
leading order, comparison of the properties of quark and gluon jets,
the comparison of the pattern of QCD radiation in different hard
scattering processes, determinations of the strong coupling, the 
study of the dynamics of the underlying subprocesses and the study of
the pattern of parton radiation. Therefore, jet substructure provides
a powerful tool to make stringent tests of perturbative QCD.

\vspace{0.5cm}
\noindent {\bf Acknowledgments}.
I would like to thank the organisers for giving me the
opportunity of presenting these results and for a well organised
conference.


\begin{thebibliography}{9}

\bibitem{prl:69:3615} 
S.D. Ellis, Z. Kunszt and D.E. Soper, \Journal{\PRL}{69}{1992}{3615}.

\bibitem{np:b383:419}
S. Catani et al., \Journal{\NPB}{383}{1992}{419}.

\bibitem{pr:d45:1448}
CDF Collab., F. Abe et al., \Journal{\PRD}{45}{1992}{1448}.

\bibitem{np:b406:187}
S. Catani et al., \Journal{\NPB}{406}{1993}{187}.

\bibitem{cpc:46:43}
H.-U. Bengtsson and T. Sj\"ostrand, \Journal{\CPC}{46}{1987}{43};
  T. Sj\"ostrand, \Journal{\CPC}{82}{1994}{74}.

\bibitem{cpc:67:465}
G. Marchesini et al., \Journal{\CPC}{67}{1992}{465}.

\bibitem{cpc:71:15}
L. Lonnblad, \Journal{\CPC}{71}{1992}{15}.

\bibitem{cpc:101:108}
G. Ingelman, A. Edin and J. Rathsman \Journal{\CPC}{101}{1997}{108}.

\bibitem{epj:c2:61}
ZEUS Collab., J. Breitweg et al., \Journal{\EPC}{2}{1998}{61}.

\bibitem{np:b700:3}
ZEUS Collab., S. Chekanov et al., \Journal{\NPB}{700}{2004}{3}.

\bibitem{h1-prelim-05-077}
H1 Collab., H1-prelim-05-077, 2005.

\bibitem{np:b545:3}
H1 Collab., C. Adloff et al., \Journal{\NPB}{545}{1999}{3}.

\bibitem{np:b485:291}
S. Catani and M.H. Seymour, \Journal{\NPB}{485}{1997}{291}.

\bibitem{pl:b558:41}
ZEUS Collab., S. Chekanov et al., \Journal{\PLB}{558}{2003}{41}.

\bibitem{epj:c8:367}
ZEUS Collab., J. Breitweg et al., \Journal{\EPC}{8}{1999}{367}.

\bibitem{epj:c31:149}
ZEUS Collab., S. Chekanov et al., \Journal{\EPC}{31}{2003}{149}.

\bibitem{pl:b380:205}
E. Mirkes and D. Zeppenfeld, \Journal{\PLB}{380}{1996}{205}.

\bibitem{prl:70:713} 
CDF Collab., F. Abe et al., \Journal{\PRL}{70}{1993}{713}.

\bibitem{pl:b357:500}
D\O\ Collab., S. Abachi et al., \Journal{\PLB}{357}{1995}{500}.

\bibitem{zfp:c63:197}
OPAL Collab., R. Akers et al., \Journal{\ZPC}{63}{1994}{197}.

\bibitem{zeus:prelim:2001}
ZEUS Collab., contributed paper to ``International Europhysics
Conference on High Energy Physics'', Budapest, Hungary, July 2001.

\bibitem{zeus-prel-07-013}
ZEUS Collab., ZEUS-prel-07-013, 2007.

\bibitem{prl:87:082001}
Z. Nagy and Z. Trocsanyi, \Journal{\PRL}{87}{2001}{082001}.

\bibitem{zeus-pub-08-014}
ZEUS Collab., S. Chekanov et al., DESY 08-178, December 2008.


\end{thebibliography}
\end{document}